\documentclass[aps,pre,twocolumn,notitlepage,superscriptaddress,floatfix,longbibliography]{revtex4-1} 
\usepackage{amsmath,amssymb}
\usepackage{xcolor}
\usepackage{graphicx}
\usepackage[unicode=true,bookmarks=false,breaklinks=false,pdfborder={0 0 1},colorlinks=true]{hyperref}
\hypersetup{urlcolor=darkblue,citecolor=darkblue,linkcolor=darkred}

\definecolor{darkblue}{rgb}{0,0,0.6}
\definecolor{darkred}{rgb}{0.6,0,0}

\begin{document}

\title{Exact solutions for the wrinkle patterns of confined elastic shells}
\author{Ian Tobasco}
 \email{itobasco@uic.edu}
 \affiliation{Department of Mathematics, Statistics, and Computer Science, University of Illinois at Chicago, Chicago, IL 60607, USA}
\author{Yousra Timounay}
 \affiliation{Department of Physics, Syracuse University, Syracuse, NY 13244, USA}
 \affiliation{BioInspired Syracuse: Institute for Material and Living Systems, Syracuse University, Syracuse, NY 13244, USA}
\author{Desislava Todorova}
 \affiliation{Department of Physics and Astronomy, University of Pennsylvania, Philadelphia, PA 19104, USA}
\author{Graham C. Leggat}
 \affiliation{Department of Physics, Syracuse University, Syracuse, NY 13244, USA}
\author{Joseph D. Paulsen}
 \email{jdpaulse@syr.edu}
 \affiliation{Department of Physics, Syracuse University, Syracuse, NY 13244, USA}
 \affiliation{BioInspired Syracuse: Institute for Material and Living Systems, Syracuse University, Syracuse, NY 13244, USA}
\author{Eleni Katifori}
 \email{katifori@sas.upenn.edu}
 \affiliation{Department of Physics and Astronomy, University of Pennsylvania, Philadelphia, PA 19104, USA}

\begin{abstract} 
Complex textured surfaces occur in nature and industry, from fingerprints to lithography-based micropatterns. Wrinkling by confinement to an incompatible substrate is an attractive way of generating reconfigurable patterned topographies, but controlling the often asymmetric and apparently stochastic wrinkles that result remains an elusive goal. Here, we describe a new approach to understanding the wrinkles of confined elastic shells, using a Lagrange multiplier in place of stress. Our theory reveals a simple set of geometric rules predicting the emergence and layout of orderly wrinkles, and explaining a surprisingly generic co-existence of ordered and disordered wrinkle domains. The results agree with numerous test cases across simulation and experiment and represent an elementary geometric toolkit for designing complex wrinkle patterns.

\end{abstract}

\maketitle


\noindent {\it Introduction.}
Dried fruits wrinkle for the same reason that leaves and flowers do --- mechanical instabilities arising from a mismatch in lengths \cite{Sharon02, Cerda03, audoly2010from, shyer2013villification, Gemmer2016, Xu2020water, Fei2020}. A similar mismatch  manifests when a thin elastic shell adheres to a substrate of a different shape \cite{Hure12, King12, Paulsen19, Vella2019buffering, Timounay21}. Can such incompatibilities be used to design and control  complex wrinkled surfaces at will? Wrinkles have been in the limelight for their theoretical importance in understanding geometric nonlinearities in elasticity \cite{Breid13, Stoop15, Reis2015, Aharoni17, bella2017wrinkling, zhang2019nonuniform, davidovitch2019, Tovkach2020}, and also for their practical significance in emerging engineering applications such as lithography-free micropatterning \cite{pretzl2008lithography, Yang2010, Chen2012, Li2017}. Yet, despite decades of study, a general predictive theory of confinement-induced wrinkling remains elusive. Such a theory would enable the creation of targeted yet reconfigurable wrinkle patterns, and could identify the broadest possible class of wrinkle morphologies that can be obtained through geometrically-incompatible confinement.

\begin{figure*}
\begin{center} 
\includegraphics[width=0.92\linewidth]{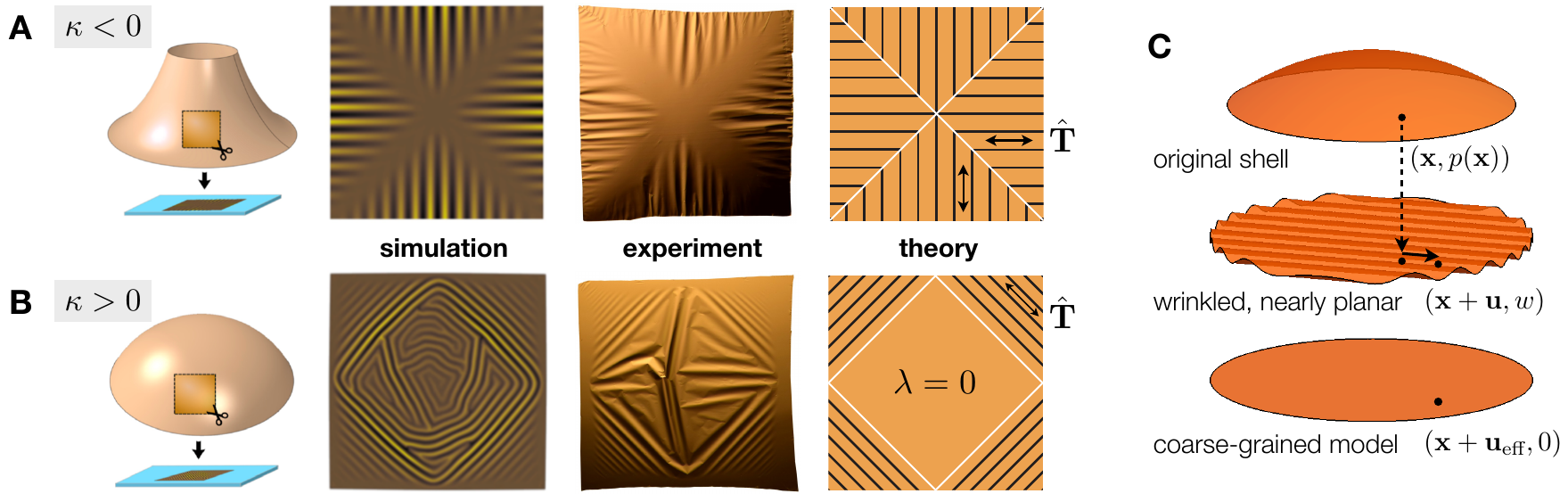}
\end{center}
\caption{
\textbf{Wrinkling of confined shells.} 
Wrinkle patterns result when initially curved shells  are confined nearby a plane. 
Simulations and experiments of square cutouts from a saddle (A) and a sphere (B) show domains of robustly ordered wrinkles, alongside a more disorderly response in the spherical case (the central diamonds in (B)). We present a coarse-grained theory to predict the type and layout of such wrinkle domains.  
(C): Coarse-graining wrinkles. A point $(\mathbf{x},p(\mathbf{x}))$ in the initial shell is displaced along the plane by $\mathbf{u}$ and out of the plane to a height $w$. The coarse-grained fields  $\mathbf{u}_\text{eff}$ and $w_\text{eff}=0$ express a theoretical limit in which the shell is infinitesimally wrinkled and perfectly confined.  
}
\label{fig:setup}
\end{figure*}

Predicting the wrinkling of confined elastic shells is a difficult problem of nonlinear mechanics. Basic theoretical issues stem from a lack of applied tensile forces that would act to organize the response. In problems dominated by tension, the guiding principle is known as tension field theory \cite{wagner1929ebene, Pipkin86, Steigmann90}, and solving it is the first step in the far-from-threshold expansion that has explained many tension-driven  patterns  \cite{Davidovitch11, King12, BellaKohn2014, Hohlfeld15, Vella15, Taffetani17}. Organized wrinkles nevertheless manifest in confined shells subject to weak or even zero tensile loads \cite{Hure12, Aharoni17}, raising the question of what sets their features. Though theoretical methods beyond tension field theory have been devised \cite{davidovitch2019}, their use requires advance knowledge of the wrinkled topography. Here, we show using theory, experiment, and simulation that the wrinkles of confined shells are in fact predicted by a compact set of simple, geometric rules. We derive our rules using a stress-like Lagrange multiplier, that arises from a Maximum Coverage Problem for the macroscopic displacement of the shell (Eq.~\eqref{eq:min-A-pblm}).

These rules imply a string of predictions about the nature of confinement-driven wrinkling, which we confirm using experiments and simulations over a broad range of parameters and shell shapes. 
As we prove, a typical shell exhibits finitely many ordered wrinkle domains where the wrinkle layout is robust; the theory also anticipates the existence of disordered wrinkle domains, whose local features behave stochastically but whose location is well-defined (Fig.~\ref{fig:setup}a,b). Second, the arrangement of these domains, and their division into ordered versus disordered, is fundamentally tied to the  shell's medial axis, a distinguished locus of points from geometry. Third, although the wrinkle amplitude  depends on the details of the shell's natural Gaussian curvature, within an ordered domain the wrinkle topography actually depends only on its sign. Finally, and perhaps most surprisingly, the wrinkle domains of  oppositely curved shells are reciprocally related, so that the response of a given shell can be deduced from another. 
Although our study focuses on the model problem of a shallow shell confined to a plane, we imagine a similar approach can be taken to understand confinement-driven patterns more generally, including ones arising from differential growth or in response to external stimuli \cite{Amar13,Reis2015,vanRees17}. We turn to introduce the setup of our study and to state our rules.

\begin{figure*}
\begin{center} 
\includegraphics[width=\linewidth]{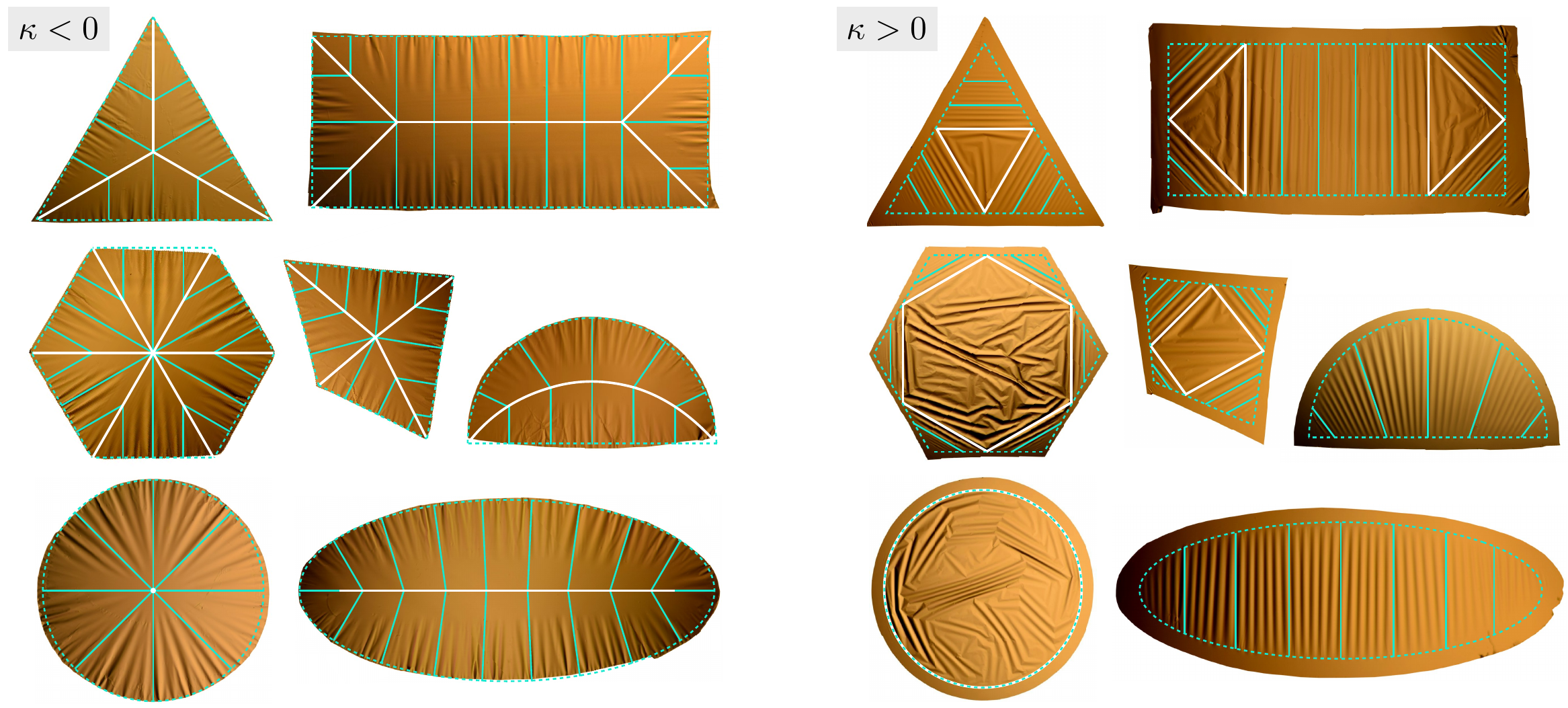}
\end{center}
\caption{
\textbf{Floating shells.}
Solid cyan lines show the directions of the field $\hat{\mathbf{T}}$ determined by solving the coarse-grained theory, overlaid on floating shells arranged by the sign of their initial Gaussian curvature (saddle-shaped on the left, spherical on the right). Regions covered by these lines are predicted to be ordered ($\lambda>0$ in the theory). Any disorder is predicted to occur in regions absent these lines (where $\lambda=0$). For saddle shells, wrinkles decay towards the medial axis in white. For spherical shells, wrinkles decay towards the boundary. Dotted cyan curves show ideal shapes used in the predictions; flattened regions are treated in the theory as infinitesimally fine. 
Experimental parameters are in Supplementary Tables S1-S2. 
}
\label{fig:stable-lines}
\end{figure*}

\smallskip 
\noindent {\it Confined shells.} A prototypical setup for confinement-driven wrinkling is shown in  Fig.~\ref{fig:setup}a,b, where square domains are cut out from a thin saddle shell or spherical cap, and are  confined to an initially planar liquid bath. 
By Gauss's Theorema Egregium, no length-preserving map exists from a curved surface to the plane. Here, this geometric incompatibility manifests as a mechanical instability producing a wrinkle pattern. Figure~\ref{fig:stable-lines} shows similar wrinkles obtained by altering the cutout shape from a square to a triangle, rectangle, ellipse, or some other shape altogether. The layout of the resulting patterns depends strongly on the chosen cutout shape, as well as on the sign of the shell's initial Gaussian curvature $\kappa$, which is negative for saddle cutouts and positive for spherical ones. Complicating things further, the typical spherical shell exhibits a mixed ``ordered--disordered'' response: in disordered regions such as the central diamonds in Fig.~\ref{fig:setup}b, the response is sensitive to perturbations and changes between trials; in ordered regions, the wrinkles are robust and repeatable.

\begin{figure}[b]
\begin{center} 
\includegraphics[width=\linewidth]{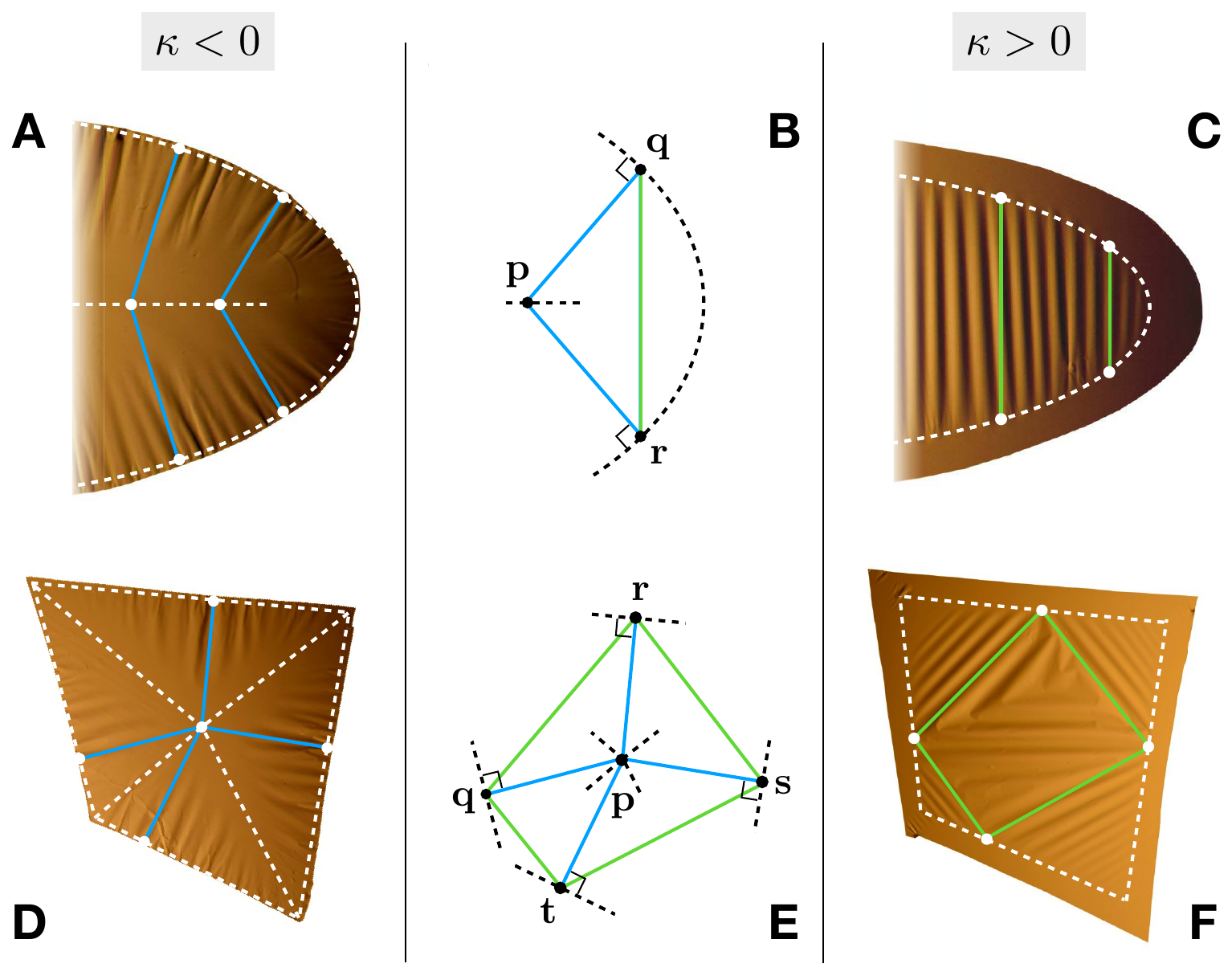}
\end{center}
\caption{\textbf{Simple rules for wrinkles.} (A-C): Ordered wrinkles in initially saddle and spherical shells (right halves of elliptical cutouts shown). Ordered wrinkles pair between shells to form a family of isosceles triangles  (B) determined by the theory. 
For saddle shells (A), wrinkles follow the blue segments $\mathbf{p}\mathbf{q}$, $\mathbf{p}\mathbf{r}$; for spherical shells (C), they follow the green segment $\mathbf{q}\mathbf{r}$. The point $\mathbf{p}$ is on the medial axis, and $\mathbf{q},\mathbf{r}$ are its closest boundary points. (D-F): Finding a disordered domain. In the given cutout shape, the point $\mathbf{p}$ has more than two closest boundary points (here, $\mathbf{q},\mathbf{r},\mathbf{s},\mathbf{t}$).  
For the saddle shell (D), ordered wrinkles follow the blue segments $\mathbf{p}\mathbf{q}$, $\mathbf{p}\mathbf{r}$, $\mathbf{p}\mathbf{s}$, $\mathbf{p}\mathbf{t}$. For the spherical shell (F), the green polygon $\mathbf{q}\mathbf{r}\mathbf{s}\mathbf{t}$ is disordered.}
\label{fig:reciprocity}
\end{figure}

 To decipher this zoo of patterns, look first at the wrinkles of the  saddle cutouts in Figs.~\ref{fig:setup}-\ref{fig:stable-lines} ($\kappa<0$).
Apparently, their wrinkles fall along  \textit{paths of quickest exit} from the cut out shape. Such paths are line segments that meet the boundary perpendicularly, and meet each other at the \textit{medial axis} or \textit{skeleton} of the shell --- the locus of points equidistant by closest approach to multiple boundary points, in white.  
Now look at the spherical cutouts ($\kappa>0$). Their wrinkles are also set by the medial axis, although this fact is not  immediately clear. 
The key is Fig.~\ref{fig:reciprocity}, which reveals that the wrinkles of saddle and spherical shells come in \textit{reciprocal pairs}. 
Most points $\mathbf{p}$ on the medial axis have exactly two closest boundary points, called $\mathbf{q}$ and $\mathbf{r}$ in panel (b). While saddle cutouts wrinkle along the segments $\mathbf{p}\mathbf{q}$ and $\mathbf{p}\mathbf{r}$, spherical cutouts wrinkle along the segment $\mathbf{q}\mathbf{r}$. Taken together, the ordered wrinkles of saddle and spherical shells form the legs of a special family of isosceles triangles whose layout is determined by the medial axis as shown. 

Notably, the legs of these isosceles triangles do not always cover the entire shell: there can exist ``leftover'' regions linked to exceptional points $\mathbf{p}$ on the medial axis with three or more closest boundary points. Figure~\ref{fig:reciprocity}(e) shows one such $\mathbf{p}$ and its four closest boundary points $\mathbf{q}$, $\mathbf{r}$, $\mathbf{s}$, and $\mathbf{t}$. While $\mathbf{p}\mathbf{q}$, $\mathbf{p}\mathbf{r}$, $\mathbf{p}\mathbf{s}$, and $\mathbf{p}\mathbf{t}$ are along the ordered wrinkles of the saddle cutout, the polygon $\mathbf{q}\mathbf{r}\mathbf{s}\mathbf{t}$ supports disorder for its spherical twin. In general, the convex hull of three or more closest boundary points can support disorder in a spherical cutout. 
The possibility of infinitely many closest boundary points occurs for a spherical disc: it is totally disordered in our simulations and experiments, save for a small flattened rim \cite{Timounay21}.

These simple rules successfully capture wrinkle patterns across $111$ experiments and several hundred more simulations. In the experiment, polystyrene films (Young's modulus $E = 3.4$ GPa, Poisson's ratio $\nu=0.34$) of thickness $120 < t < 430$~nm are spin-coated on curved glass substrates. 
The spherical or saddle shape of the substrate imparts a finite rest curvature on the shell, with principal radii of curvature $R$ ranging from $13$ to $39$~mm. 
Cutouts of width $2.5 < W < 16$~mm are released onto a flat water bath with surface tension $\gamma_{lv}=0.072$~N/m and gravitational stiffness $K=\rho g$. 
The experiments reside in the limit of weak tension $\gamma_{lv} R^2 \ll Y W^2$, 
moderately stiff substrate $K W^2 \gtrsim \gamma_{lv}$, and small bending stiffness $B K R^4 \ll Y^2 W^4$ 
where $Y = Et$ and $B = Et^3/[12 (1-\nu^2)]$ are the stretching and bending moduli. 
Being shallow yet much larger than the characteristic substrate-dominated wrinkle wavelength, $(B/K)^{1/4} \ll W \ll R$,
the cutouts adopt approximately planar shapes and wrinkle as they float on the water bath, as shown in  Figs.~\ref{fig:setup}-\ref{fig:stable-lines}. 

\begin{figure}
\begin{center}
\includegraphics[width=\linewidth]{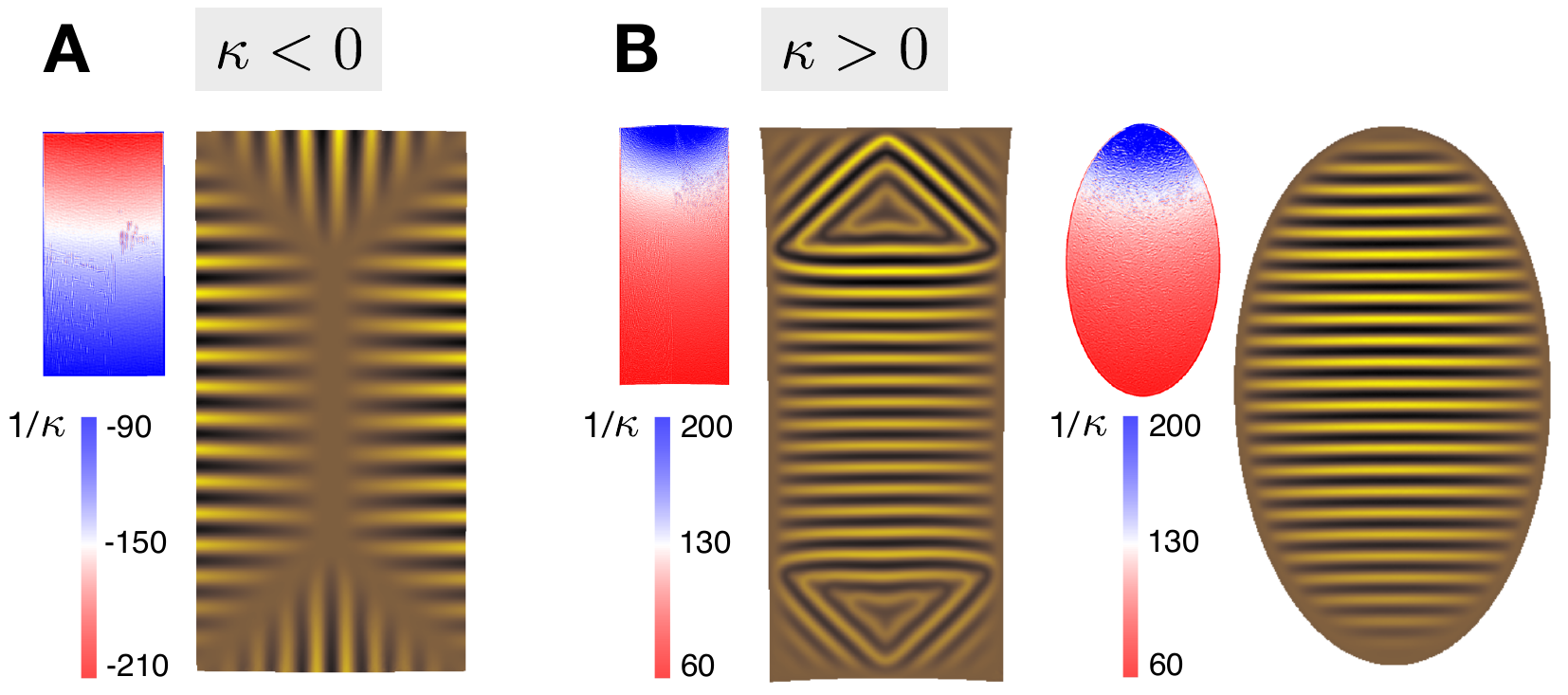}
\caption{\textbf{Variable-curvature shells.} Simulations show initially negatively (A) or positively curved (B) shells confined nearby a plane. Colormaps show the inverse of the initial Gaussian curvature
$\kappa(\mathbf{x})$. The patterns are  independent of its precise values, but depend strongly on its sign. The wrinkle amplitude reflects the curvature.
Simulation parameters are in Supplementary Table S3.
}
\label{fig:nonconstant} 
\end{center}
\end{figure}

To probe the role of surface tension in setting the patterns, we perform ABAQUS simulations of shells on a liquid substrate in a similar parameter regime, but with the surface tension set to zero so that no forces are applied at the lateral shell boundary. The result is a ``softly stamped'' version of the well-known example of a plate pressed into a hard spherical mold  \cite{Hure12,davidovitch2019}. Similar patterns arise in the simulation and the experiment, with the layout of the ordered domains and the location of the disordered domains being the same (Fig.~\ref{fig:setup}a,b). The conclusion, which should be compared against the paradigm of tension as an organizing force determining wrinkle patterns \cite{Pipkin86,Steigmann90}, is that well-defined and spatially-complex patterns persist \textit{even without applied tensile forces}. 
Simulations of shells with non-constant initial Gaussian curvatures $\kappa(\mathbf{x})$ reveal an even more curious fact: so long as the initial Gaussian curvature of a shell is of one sign --- either positive or negative everywhere --- the  patterns are the same as for shells with approximately constant curvature (compare Figs.~\ref{fig:stable-lines} and \ref{fig:nonconstant}).

\smallskip
\noindent {\it Minimizing energy by maximizing coverage.} We turn to explain these remarkably robust features of confinement-driven wrinkling, and to derive our simple rules. We do so by analyzing a novel coarse-grained model for incompatibly confined shallow shells from Ref.~\cite{tobasco2021curvature}, which we summarize now.
Consider the setup in Fig.~\ref{fig:setup}c, where a material point $(\mathbf{x},p(\mathbf{x}))$ in the initial shell displaces to $(\mathbf{x}+\mathbf{u}(\mathbf{x}),w(\mathbf{x}))$ on the bath. The reference point $\mathbf{x}=(x_1,x_2)$ is in the shell's initial planform $\Omega\subset \mathbb{R}^2$, e.g., a square in Fig.~\ref{fig:setup}a,b. 
The displacements $\mathbf{u}=(u_1,u_2)$ and $w-p$ are respectively parallel and perpendicular to the initial bath. 
Patterns manifest through minimization of the system energy, $U=U_\text{shell}+U_\text{subs}$, where $U_\text{shell}$ is the energy of bending and stretching the shell, and $U_\text{subs}$ is the gravitational potential energy of the bath plus its liquid surface energy (the latter is set to zero in the simulations) {\cite{audoly2010from,ciarlet1997mathematical}}.

Energy minimizations of this type are usually solved using tension field theory \cite{Pipkin86,Steigmann90}, which involves an expansion about a uniaxially or biaxially tensile   \textit{effective displacement} $(\mathbf{u}_\text{eff},-p)$. This effective state is obtained by coarse-graining away the wrinkles from the shell's physical displacement $(\mathbf{u},w-p)$, in a limit where the wrinkle wavelength and amplitude go to zero (Fig.~\ref{fig:setup}c). The typical explanation is that the direction of the wrinkles is set by tensile boundary loads, which stabilize their peaks and troughs. 
Yet, our patterns occur in confined shells subject to small or even zero boundary loads, suggesting an alternate expansion about a uniaxially or biaxially compressive state --- one that is \textit{tension-free}. 
To motivate this further, note that in such a situation, one may expect the stretching energy of the shell to be subdominant to its bending and substrate energies \cite{davidovitch2019}. 
Under a simplifying hypothesis guaranteeing this hierarchy, Ref.~\cite{tobasco2021curvature} obtained an expansion of the system energy about a general tension-free state.
As obtained, this expansion is outside the parameter range of the experiments and simulations.
Nevertheless, for each $\mathbf{u}_\text{eff}$ the energy was found to be proportional to $\gamma_\text{eff} := \gamma_{lv} + 2\sqrt{BK}$ {with $U/\gamma_\text{eff} = \int_\Omega \frac{1}{2}|\nabla w|^2\,d\mathbf{x}$ at leading order}, up to a constant not depending on {the effective state.}

{To bring this into a more useful form, note the strain $\varepsilon_{ij}(\mathbf{u},w) =(\partial_i u_j+\partial_j u_i + \partial_i w \partial_j w -\partial_i p \partial_j p)/2$, $i,j\in\{1,2\}$ tends to zero in the expansion, so that the shell's total area is asymptotically conserved: 
\begin{equation}
\Delta A_\text{tot}= \int_\Omega  \frac{1}{2}|\nabla p|^2\,d\mathbf{x}-\int_\Omega  \nabla \cdot \mathbf{u} +\frac{1}{2}|\nabla w|^2\,d\mathbf{x} \to 0.
\end{equation}
Taking $\mathbf{u}\to\mathbf{u}_\text{eff}$ gives the following expression for the leading order energy of a confined shallow shell \cite{tobasco2021curvature}:} 
\begin{equation}\label{eq:effective-energy}
\frac{U}{\gamma_\text{eff}}\to \int_\Omega \frac12|\nabla p|^2\,d\mathbf{x} - \int_{\partial\Omega} \mathbf{u}_\text{eff}\cdot\hat{\mathbf{n}}\,ds := \Delta A\{\mathbf{u}_\text{eff}\}
\end{equation}
where $\hat{\mathbf{n}}$ is the outwards pointing unit normal to the boundary, $\partial\Omega$.  
{This way of writing the energy emphasizes the role of the difference $\Delta A = A_\text{init} - A_\text{eff}$ between the shell's initial area, $A_\text{init}=\int_\Omega 1+\frac{1}{2}|\nabla p|^2\,d\mathbf{x}$, and the area covered by its infinitesimally wrinkled, perfectly planar limit, $A_\text{eff}=\int_\Omega 1+\nabla\cdot\mathbf{u}_\text{eff}\,d\mathbf{x}$. This difference accounts for the area that is ``lost'' asymptotically to wrinkles. It sets the energy of confinement per Eq.~\eqref{eq:effective-energy}.}
Notably, this energy is not proportional to the stretching modulus $Y$ as significant tension is not involved. Indeed, \eqref{eq:effective-energy} was found to hold even in situations lacking boundary loads ($\gamma_{lv}=0$)  --- a stark difference from tension field theory. And as we will show, the analysis of Eq.~\eqref{eq:effective-energy} leads to the patterns observed in our experiments and simulations, raising the question of whether it can be justified in a wider parameter range.

Optimizing the result of Eq.~\eqref{eq:effective-energy} determines the effective displacement of the shell. To help visualize this, imagine first projecting the shell directly into the plane, such that it is in a state of total compression. While this compression can be relieved by wrinkling, it can also be reduced by lateral displacements within the plane. These displacements take advantage of the liquid nature of the bath, which allows the shell to ``get out of its own way''. Their typical magnitude is $\sim W^3/R^2$, making them much larger than the wrinkles, whose lateral oscillations are $\sim (B/K)^{1/4}(W/R)^2$. The bulk lateral displacements are called $\mathbf{u}_\text{eff}$ in the coarse-grained theory and are selected to minimize the cost of their accompanying wrinkles, captured by  $\Delta A$.

Importantly, this minimization is done under the constraint that $(\mathbf{u}_\text{eff},-p)$  is tension-free, to prevent the shell from stretching at a higher energy cost. To enforce it, we use the \textit{effective strain} {$(\boldsymbol{\varepsilon}_\text{eff})_{ij}(\mathbf{u}_\text{eff}) :=(\partial_i (\mathbf{u}_\text{eff})_j+\partial_j (\mathbf{u}_\text{eff})_i -\partial_i p \partial_j p)/2$} 
gotten by setting $\mathbf{u}=\mathbf{u}_\text{eff}$ and $w=0$ into the {previous} formula for the strain of a shallow shell. 
While the physical strain $\boldsymbol{\varepsilon}$ tends to zero, the effective strain $\boldsymbol{\varepsilon}_\text{eff}$ is non-zero due to wrinkling. Its eigenvalues are constrained to be non-positive, a situation we denote by $\boldsymbol{\varepsilon}_\text{eff}\leq \mathbf{0}$. 
{As in tension field theory, a strictly negative eigenvalue indicates a length lost to wrinkles in the limit; a zero eigenvalue means that length is preserved.}
Thus, we arrive at the Maximum Coverage Problem from \cite{tobasco2021curvature}:
\begin{equation}\label{eq:min-A-pblm}
\min\,\Delta A\{\mathbf{u}_\text{eff}\} \quad\text{subject to}\quad \boldsymbol{\varepsilon}_\text{eff}(\mathbf{u}_\text{eff}) \leq \mathbf{0}.
\end{equation}
By minimizing the area lost to infinitesimal wrinkles, the coarse-grained shell covers a maximal area in the plane. Using this attractive geometric variational principle, we shall deduce the phenomenology of wrinkle domains.

\smallskip
\noindent {\it The locking stress Lagrange multiplier.}
To uncover the patterns predicted by the Maximum Coverage Problem, 
we now introduce a notion of ``effective stress'' to pair with the effective strain. Recognizing the nonholonomic nature of the constraint $\boldsymbol{\varepsilon}_\text{eff}\leq \mathbf{0}$, we replace it with a symmetric matrix-valued Lagrange multiplier field $\boldsymbol{\sigma}_L(\mathbf{x})$ we call the  \textit{locking stress} (see the Discussion for the nomenclature). We require that $\boldsymbol{\sigma}_L\geq\mathbf{0}$, meaning its eigenvalues are non-negative. We define the Lagrangian
\begin{equation}
\mathcal{L}\left\{ \mathbf{u}_\text{eff},\boldsymbol{\sigma}_{L}\right\} =\Delta A\{\mathbf{u}_\text{eff}\}+\int_{\Omega}\boldsymbol{\sigma}_{L}:\boldsymbol{\varepsilon}_\text{eff}(\mathbf{u}_\text{eff}) \label{eq:Lagrangian}
\end{equation}
for all $\mathbf{u}_\text{eff}$ and $\boldsymbol{\sigma}_L\geq\mathbf{0}$, and seek a saddle point. 
Enforcing stationarity of $\mathbf{u}_\text{eff}$, we find that  $\nabla\cdot\boldsymbol{\sigma}_L =0$ in the shell $\Omega$ and $\boldsymbol{\sigma}_L \hat{\mathbf{n}}=\hat{\mathbf{n}}$ at its boundary $\partial\Omega$. As discussed in the Methods, a relaxation of the boundary conditions ensures existence of a saddle point: we enforce them from the outside of the shell, but not necessarily from its inside. There, we also derive the orthogonality relation
\begin{equation}\label{eq:orthogonal}
\boldsymbol{\sigma}_L : \boldsymbol{\varepsilon}_\text{eff} = 0
\end{equation}
relating the locking stress to the effective strain. 
At this point, we have everything we need to solve for the wrinkle domains. Indeed, while $\boldsymbol{\sigma}_L$ is not the true stress in the shell (neither is $\boldsymbol{\varepsilon}_{\text{eff}}$ the true strain), knowledge of it reveals constraints on the patterns to the point that it is an order parameter for wrinkle domains.

\begin{figure}
\begin{center}
\includegraphics[width=0.95\linewidth]{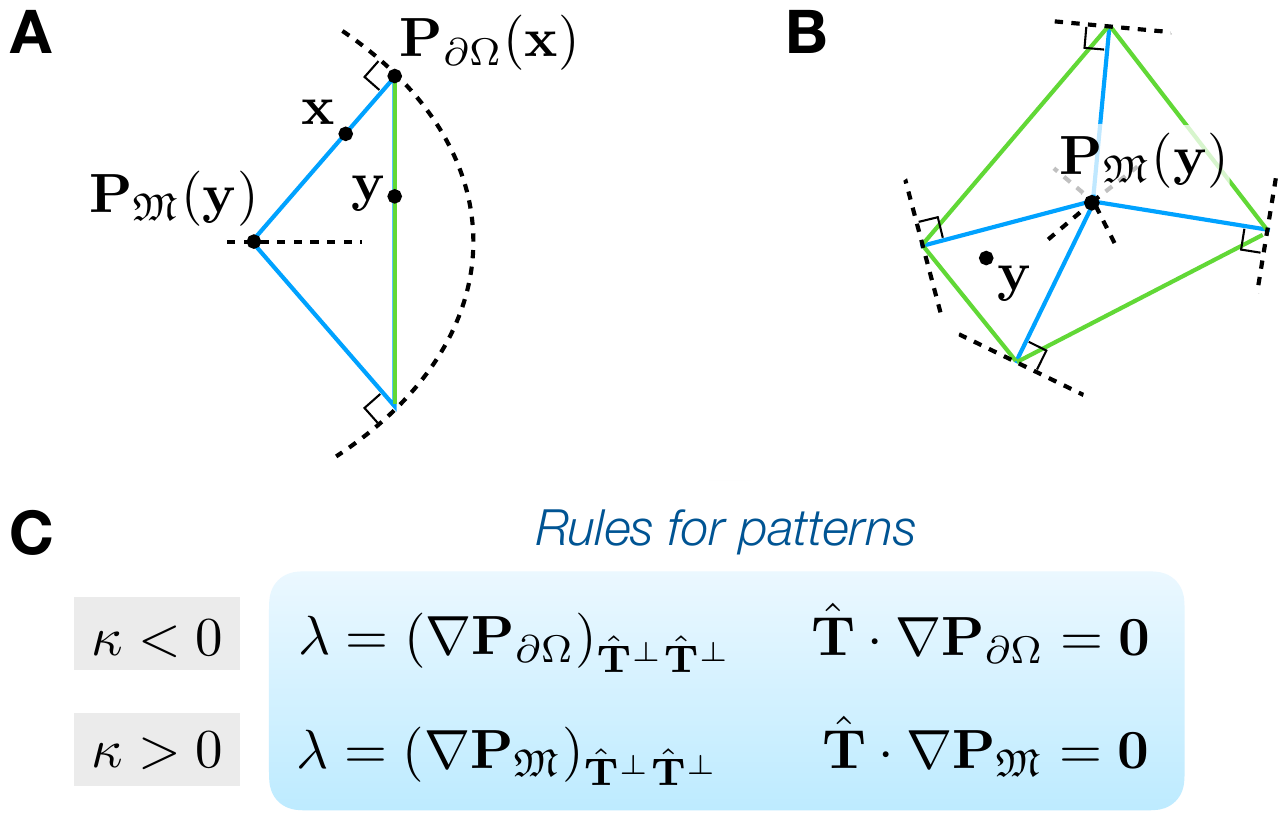}
\caption{\textbf{Deducing the simple rules.} 
(A-B): Two geometric operations take a point $\mathbf{x}$ in the shell's planform $\Omega$ to its closest boundary point $\mathbf{P}_{\partial\Omega}(\mathbf{x})$ (A), or take a point $\mathbf{y}$ in $\Omega$ to a  point $\mathbf{P}_{\mathfrak{M}}(\mathbf{y})$ on the medial axis whose closest boundary points have $\mathbf{y}$ in their convex hull (B). (C): The fields $\lambda$ and $\hat{\mathbf{T}}$  governing wrinkle patterns depend on the shell through these operations and the sign of its initial Gaussian curvature $\kappa$. 
}
\label{fig:maps} 
\end{center}
\end{figure}

To explain this last remark further, note first that $\boldsymbol{\varepsilon}_\text{eff}$ cannot vanish where the initial Gaussian curvature $\kappa(\mathbf{x})$ of the shell is non-zero.  
Due to the orthogonality in \eqref{eq:orthogonal},  
\begin{equation}\label{eq:rank-one}
\boldsymbol{\sigma}_L = \lambda \hat{\mathbf{T}}\otimes\hat{\mathbf{T}}
\end{equation}
for some scalar and unit vector fields $\lambda(\mathbf{x})\geq 0$ and $\hat{\mathbf{T}}(\mathbf{x})$ which contain information about the patterns. 
In particular, by \eqref{eq:orthogonal} and \eqref{eq:rank-one}, these quantities obey 
\begin{equation}\label{eq:important-identity}
\lambda \cdot (\boldsymbol{\varepsilon}_\text{eff})_{\hat{\mathbf{T}}\hat{\mathbf{T}}}=0.
\end{equation}
In regions where $\lambda>0$ the $\hat{\mathbf{T}}\hat{\mathbf{T}}$-component of $\boldsymbol{\varepsilon}_\text{eff}$ must vanish, indicating an ordered domain with wrinkle peaks and troughs along $\hat{\mathbf{T}}$. Conversely, where $\lambda =0$ the wrinkle direction is unconstrained, permitting a disordered response. The type and layout of a given shell's wrinkle domains are predicted by the locking stress.

Remarkably, it is possible to find the locking stress of a  shell without first determining its effective strain, an observation that leads to a complete derivation of our simple rules.
Eliminating $\mathbf{u}_\text{eff}$ from  the Lagrangian in \eqref{eq:Lagrangian} by minimization yields a separate, ``dual'' variational principle for $\boldsymbol{\sigma}_L$ (Methods Eq.~\eqref{eq:dual-pblm}). We solve it exactly in the Supplementary Information, using convex Airy potentials and an inspired application of the Legendre transform. The resulting solution formulas determine $\lambda$ and $\hat{\mathbf{T}}$ by one of two basic geometric operations, shown in  Fig.~\ref{fig:maps}. These formulas apply whenever the shell has no holes and if its initial Gaussian curvature $\kappa(\mathbf{x})$ is of one sign. They are the basis of our simple rules (compare Figs.~\ref{fig:reciprocity} and \ref{fig:maps}). For instance, the fact that $\hat{\mathbf{T}}\cdot \nabla \mathbf{P}_{\partial\Omega}=\mathbf{0}$ if $\kappa <0$ explains why the wrinkles of negatively curved shells lie along directions of quickest exit to the shell boundary. The remaining rules are derived in the SI. 

Coming back to our experiments and simulations, we now derive their patterns. In Fig.~\ref{fig:stable-lines}, cyan lines are drawn along the solved-for $\hat{\mathbf{T}}$ in the predicted ordered regions (where $\lambda>0$). Each negatively curved shell is found to be completely ordered. Regions consistent with disorder (where $\lambda=0$) exist for generic positively curved shells, and are shown as polygons bordered in white. The wrinkle domains are set by the shells' medial axes following our simple rules.

\begin{figure}
\begin{center} 
\includegraphics[width=.9\linewidth]{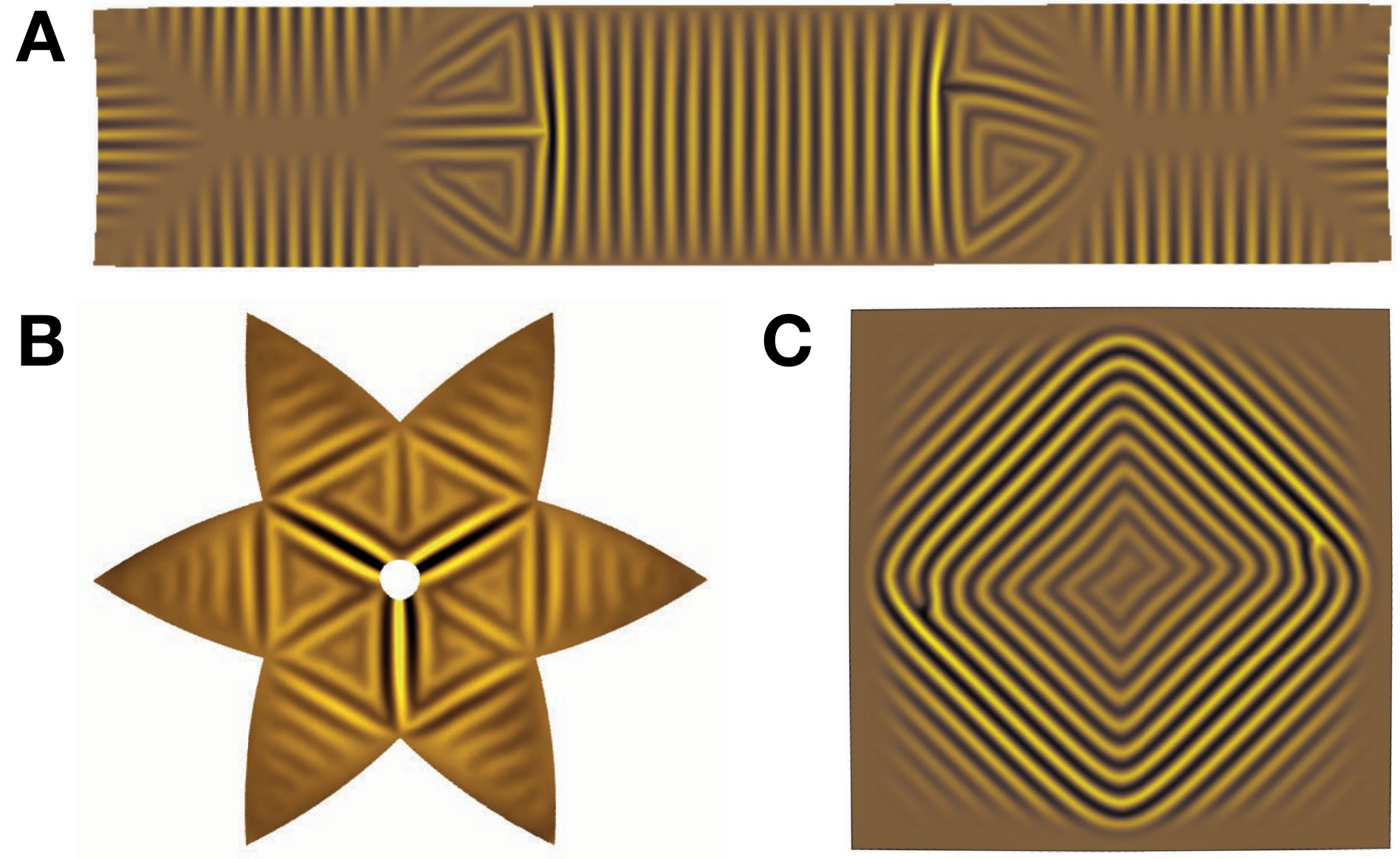}
\caption{\textbf{Open questions.} 
While the locking stress can be defined for  (A) shells with mixed curvature ($\kappa <0$ on the left and right, $\kappa >0$ in the middle) and (B) shells with holes ($\kappa>0$ here), we lack solution formulas for it in such cases. (C) Another question regards the presence of order in regions consistent with disorder (compare with Fig.~\ref{fig:setup}b). 
}
\label{fig:questions}
\end{center}
\end{figure}

\smallskip
\noindent {\it Discussion.}
Given the success of our rules in capturing the wrinkles of confined shells, it is natural to consider other instances of reciprocity as well as graphical methods in mechanics more broadly. A well-known method is due to Maxwell \cite{maxwell1864reciprocal}  
and also Taylor, whose reciprocal diagrams of forces and frames encode an elegant test of equilibrium for planar structures \cite{timoshenko1953history}. 
Our relations connecting the wrinkles of positively and negatively curved shells reveal a new class of reciprocal rules governing incompatible confinement. 
We wonder how far they generalize. Examples of shells for which we presently lack rules are in Fig.~\ref{fig:questions}a,b. Finally, Fig.~\ref{fig:questions}c highlights the fact that ordered wrinkles sometimes occur in regions the theory predicts to be consistent with disorder. 
Empirically, the presence of order versus disorder looks to depend on the finite wrinkle wavelength. 
Related to this is the question of the greatest parameter regime in which the Maximum Coverage Problem in Eq.~\eqref{eq:min-A-pblm} can be derived. Though it predicts the patterns in our simulations and experiments well, it has yet to be established for the parameter range they explore. We imagine a full proof of Eq.~\eqref{eq:min-A-pblm} will come from combining the ``inverted tension field theory'' of \cite{davidovitch2019} with the ansatz-free arguments in \cite{tobasco2021curvature}. 

We have shown how to predict the wrinkles of confined shallow shells, using a compact set of geometric rules gotten by solving the coarse-grained theory of \cite{tobasco2021curvature}. Our results point towards a general, diagrammatic method for benchmarking elastic patterns, which could prove useful for their rapid design.
We highlight a promising connection with the theory of \textit{ideal locking materials} --- bulk materials whose microstructures facilitate extension with negligible elastic stress below a threshold strain \cite{Prager1957}. This limit is apparently approached in biology, by the mesentery membrane of rabbits  \cite{fung1967elasticity,Prager1969} and the capture silk of some spiders, the latter of which has recently inspired ultrastretchable wicked membranes \cite{elettro2016drop,grandgeorge2018capillarity}. 
We view the wrinkles of confined shells as an emergent-yet-sacrificial microstructure enabling shape change. This underlies our terming the Lagrange multiplier $\boldsymbol{\sigma}_L$ from our solutions as the \textit{locking stress}. 
It plays the role of an order parameter for predicting wrinkle domains.
The extension of our rules beyond shallow shells and to patterns involving elements others than wrinkles, including crumples \cite{King12,Timounay20} and folds \cite{Pocivavsek08,Brau13,Paulsen17}, remains to be seen.

\smallskip
\noindent {\it Methods}.

\smallskip
\noindent {\it Experiment.} Dilute solutions of polystyrene ($M_\text{n} = 99$ kDa, $M_\text{w} = 105.5$ kDa, Polymer Source) in toluene (99.9\%, Fisher Scientific) are spin-coated onto glass substrates
of various positive and negative Gaussian curvatures. 
The positively curved substrates are spherical optical lenses (Thorlabs, Inc.). Negatively curved  shells were formed on a single negative-curvature substrate that is less controlled by comparison; its principal radii of curvature were measured from side-view images and are reported in Supplementary Table S1.

Film thickness is varied by changing the polymer concentration
and spinning speed. 
Different shapes are cut out using a metal scribe. After preparing
the glass substrates with a thin layer of poly(acrylic acid), the
films are released by dissolving this sacrificial layer in water.
The films are finally transferred to a pure water--air interface.
 Following the experiments, each film is captured and its thickness is measured
using a white-light interferometer (Filmetrics F3). 

The shells are shallow with $0.01<(W/R)^{2}<0.2$ and have non-dimensional bending modulus $b=BR^{2}/YW^{4}$ in the range $4\times10^{-11}<b<2\times10^{-8}$.  
The non-dimensional substrate stiffness $k=KR^{2}/Y$ and non-dimensional surface tension $\gamma=\gamma_\text{lv}R^{2}/YW^{2}$ obey $0.003<k<0.03$ and 
$4\times10^{-4}<\gamma<10^{-2}$.
Additionally, $10^{-2}<\gamma/k<0.7$, $10^{-6}<2\sqrt{bk}<3\times10^{-5}$, and $10^{-3}<\left(b/k\right)^{1/4}<10^{-1}$. 
These ranges are in line with all but one of the assumptions used in Ref.~\cite{tobasco2021curvature} to derive Eq.~\eqref{eq:effective-energy} (they do not obey $(b/k)^{1/10}\ll\gamma + 2 \sqrt{bk}$). 
Specific parameters for the experiments shown in the main text are in Supplementary Tables S1-S2. 


\smallskip
\noindent {\it Simulation.} Shells bonded to a planar liquid substrate without surface tension are simulated 
in the finite element package ABAQUS/Explicit. Four-node thin shell elements with reduced integration
(element type S4R) are used. The confining force is specified as a non-uniform distributed pressure load over the surface of the shell, via a VDLOAD subroutine. Otherwise, free boundary conditions are used. Comparative non-linear geometric finite element analysis using linearly elastic and neo-Hookean
hyperelastic materials show the results are largely independent of the model. 
Color-coding in the images corresponds to vertical deflections from the plane.

In the same non-dimensional groups as before, the simulations have $0.01<(W/R)^{2}<0.04$,  $7\times10^{-9}<b<2\times10^{-6}$, $6<k<40$, and $\gamma =0$. 
Additionally, $2\times10^{-4}<2\sqrt{bk}<2\times10^{-2}$ and $5.6\times10^{-3}<\left(b/k\right)^{1/4}<10^{-2}$. 
As with the experiments, these ranges are in line with all but one of the assumptions of Ref.~\cite{tobasco2021curvature} (they do not obey $(b/k)^{1/10}\ll 2\sqrt{bk}$). 
Specific parameters for the simulations shown in the main text are in Supplementary Table S3. 

\smallskip
\noindent {\it Theory.} Here we connect the Lagrangian $\mathcal{L}$ in Eq.~\eqref{eq:Lagrangian} of the main text to our coarse-grained fields.  
We assert the existence of a \textit{saddle point} $(\mathbf{u}_\text{eff},\boldsymbol{\sigma}_L)$ satisfying
\begin{equation}\label{eq:saddle-defn}
    \mathcal{L}\{\mathbf{u}_\text{eff}+\delta\mathbf{u}_\text{eff},\boldsymbol{\sigma}_L\}
    \geq \mathcal{L}\{\mathbf{u}_\text{eff},\boldsymbol{\sigma}_L\}\geq \mathcal{L}\{\mathbf{u}_\text{eff},\boldsymbol{\sigma}_L+\delta\boldsymbol{\sigma}_L\}
\end{equation}
for all $\delta\mathbf{u}_\text{eff}$ and $\delta\boldsymbol{\sigma}_L$ with  $\boldsymbol{\sigma}_L+\delta\boldsymbol{\sigma}_L\geq\mathbf{0}$ \cite{tobasco2021curvature}.
The key linking saddle points to the Maximum Coverage Problem  (Eq.~\eqref{eq:min-A-pblm}) is that such points yield its solutions. 
To study saddles in detail, we evaluate the ``min-max'' and ``max-min'' procedures $\min_{\mathbf{u}_\text{eff}}\max_{\boldsymbol{\sigma}_L}\mathcal{L}$ and $\max_{\boldsymbol{\sigma}_L}\min_{\mathbf{u}_\text{eff}}\mathcal{L}$. 

First, consider the min-max: we claim that  \begin{equation}\label{eq:min-max}
\min_{\mathbf{u}_\text{eff}}\max_{\boldsymbol{\sigma}_L}\,\mathcal{L}=\min_{\mathbf{u}_\text{eff}}\,\Delta A 
\end{equation}
where on the right the tension-free constraint $\boldsymbol{\varepsilon}_\text{eff}\leq\mathbf{0}$ is used. Eq.~\eqref{eq:min-max} states that solving the Maximum Coverage Problem is equivalent to finding the min-max of $\mathcal{L}$, and explains why its saddle points contain our effective displacements. 
To prove it, note the inner maximization over  $\boldsymbol{\sigma}_L\geq \mathbf{0}$ enforces the tension-free constraint: if a component of $\boldsymbol{\varepsilon}_\text{eff}$ is positive, then by sending the same component of $\boldsymbol{\sigma}_L$ to infinity we get  $\max\,\mathcal{L}=\infty$; conversely, if $\mathbf{u}_\text{eff}$ is tension-free then $\int_\Omega \boldsymbol{\sigma}_L:\boldsymbol{\varepsilon}_\text{eff}\leq 0$ and  $\max\,\mathcal{L}=\Delta A$. Evidently, minimizing the maximum prefers tension-free states. It follows that saddle points achieve $\mathcal{L}=\Delta A$ or, equivalently, that  $\int_\Omega \boldsymbol{\sigma}_L:\boldsymbol{\varepsilon}_\text{eff}= 0$. Since the integrand is non-positive it must vanish, proving the orthogonality relation \eqref{eq:orthogonal}.

Next, consider the max-min: a computation in the SI using the divergence theorem gives 
\begin{equation}\label{eq:dual-pblm}
\max_{\boldsymbol{\sigma}_L}\min_{\mathbf{u}_\text{eff}}\,\mathcal{L} = \max_{\boldsymbol{\sigma}_L}\,-\frac{1}{2}\int_{\mathbb{R}^2}\nabla p \otimes \nabla p:(\boldsymbol{\sigma}_{L}-\mathbf{I})
\end{equation}
where in the resulting maximization $\boldsymbol{\sigma}_L$ is constrained to be a non-negative symmetric matrix-valued field equaling the identity  $\mathbf{I}$ exterior to $\Omega$, and that is weakly divergence-free on $\mathbb{R}^2$.  This is the dual problem mentioned in the main text, and solving it gives the locking stress associated to $\Omega$ and $p$. The choice to extend $\boldsymbol{\sigma}_L$ beyond the shell relaxes its boundary conditions so that a maximizer always exists \cite{tobasco2021curvature}. The original boundary condition $\boldsymbol{\sigma}_L\hat{\mathbf{n}}=\hat{\mathbf{n}}$ can be thought of as happening outside of an infinitesimally thin boundary layer at $\partial\Omega$; the inner boundary values of $\boldsymbol{\sigma}_L$ can then be optimized. This relaxation is crucial to capturing the patterns of positively curved shells (see Fig.~\ref{fig:stable-lines}). The dual problem for $\boldsymbol{\sigma}_L$ is discussed further in the SI, where we solve it using convex Airy potentials  to derive our simple rules.

\smallskip
\noindent {\it Data Availability}.
The parameters for the shells in Figs.~\ref{fig:setup}-\ref{fig:nonconstant} are in Supplementary Tables S1-S3. Dimensionless parameter ranges for the experiments and simulations are in the Methods. Individual parameters for all experiments are also provided as a Supplementary Datafile. All other data that support the findings of this study are available from the corresponding authors upon reasonable request.

\smallskip
\noindent {\it Acknowledgements.} 
We thank B.\ Davidovitch, V.\ D\'emery, C.~R.\ Doering, G.\ Francfort, S.\ Hilgenfeldt, R.~D.\ James, R.~V.\ Kohn, N.\ Menon, P.\ Plucinsky, D.\ Vella, and A.\ Waas for  helpful discussions.
This work was supported by NSF awards DMS-1812831 and DMS-2025000 (IT); 
NSF award DMR-CAREER-1654102 (YT, JDP); 
NSF award PHY-CAREER-1554887, Univ.\ of Pennsylvania MRSEC award DMR-1720530 and CEMB award CMMI-1548571, and a Simons Foundation award 568888 (DT, EK).

\smallskip
\noindent {\it Author Contributions.}  
IT, EK, and JDP conceived and designed the research; 
IT developed and implemented the theory;
DT and EK conducted and analyzed the simulations;
YT, GCL, and JDP conducted and analyzed the experiments;
IT, YT, DT, EK, and JDP wrote the manuscript.

\smallskip
\noindent {\it Competing Interests.} 
The authors declare no competing interests.

\bibliography{ShellRefs.bib}

\begin{thebibliography}{47}%
\makeatletter
\providecommand \@ifxundefined [1]{%
 \@ifx{#1\undefined}
}%
\providecommand \@ifnum [1]{%
 \ifnum #1\expandafter \@firstoftwo
 \else \expandafter \@secondoftwo
 \fi
}%
\providecommand \@ifx [1]{%
 \ifx #1\expandafter \@firstoftwo
 \else \expandafter \@secondoftwo
 \fi
}%
\providecommand \natexlab [1]{#1}%
\providecommand \enquote  [1]{``#1''}%
\providecommand \bibnamefont  [1]{#1}%
\providecommand \bibfnamefont [1]{#1}%
\providecommand \citenamefont [1]{#1}%
\providecommand \href@noop [0]{\@secondoftwo}%
\providecommand \href [0]{\begingroup \@sanitize@url \@href}%
\providecommand \@href[1]{\@@startlink{#1}\@@href}%
\providecommand \@@href[1]{\endgroup#1\@@endlink}%
\providecommand \@sanitize@url [0]{\catcode `\\12\catcode `\$12\catcode
  `\&12\catcode `\#12\catcode `\^12\catcode `\_12\catcode `\%12\relax}%
\providecommand \@@startlink[1]{}%
\providecommand \@@endlink[0]{}%
\providecommand \url  [0]{\begingroup\@sanitize@url \@url }%
\providecommand \@url [1]{\endgroup\@href {#1}{\urlprefix }}%
\providecommand \urlprefix  [0]{URL }%
\providecommand \Eprint [0]{\href }%
\providecommand \doibase [0]{http://dx.doi.org/}%
\providecommand \selectlanguage [0]{\@gobble}%
\providecommand \bibinfo  [0]{\@secondoftwo}%
\providecommand \bibfield  [0]{\@secondoftwo}%
\providecommand \translation [1]{[#1]}%
\providecommand \BibitemOpen [0]{}%
\providecommand \bibitemStop [0]{}%
\providecommand \bibitemNoStop [0]{.\EOS\space}%
\providecommand \EOS [0]{\spacefactor3000\relax}%
\providecommand \BibitemShut  [1]{\csname bibitem#1\endcsname}%
\let\auto@bib@innerbib\@empty
\bibitem [{\citenamefont {Sharon}\ \emph {et~al.}(2002)\citenamefont {Sharon},
  \citenamefont {Roman}, \citenamefont {Marder}, \citenamefont {Shin},\ and\
  \citenamefont {Swinney}}]{Sharon02}%
  \BibitemOpen
  \bibfield  {author} {\bibinfo {author} {\bibfnamefont {E.}~\bibnamefont
  {Sharon}}, \bibinfo {author} {\bibfnamefont {B.}~\bibnamefont {Roman}},
  \bibinfo {author} {\bibfnamefont {M.}~\bibnamefont {Marder}}, \bibinfo
  {author} {\bibfnamefont {G.-S.}\ \bibnamefont {Shin}}, \ and\ \bibinfo
  {author} {\bibfnamefont {H.~L.}\ \bibnamefont {Swinney}},\ }\bibfield
  {title} {\enquote {\bibinfo {title} {Buckling cascades in free sheets},}\
  }\href {\doibase 10.1038/419579a} {\bibfield  {journal} {\bibinfo  {journal}
  {Nature}\ }\textbf {\bibinfo {volume} {419}},\ \bibinfo {pages} {579--579}
  (\bibinfo {year} {2002})}\BibitemShut {NoStop}%
\bibitem [{\citenamefont {Cerda}\ and\ \citenamefont
  {Mahadevan}(2003)}]{Cerda03}%
  \BibitemOpen
  \bibfield  {author} {\bibinfo {author} {\bibfnamefont {E.}~\bibnamefont
  {Cerda}}\ and\ \bibinfo {author} {\bibfnamefont {L.}~\bibnamefont
  {Mahadevan}},\ }\bibfield  {title} {\enquote {\bibinfo {title} {Geometry and
  physics of wrinkling},}\ }\href {\doibase 10.1103/PhysRevLett.90.074302}
  {\bibfield  {journal} {\bibinfo  {journal} {Phys. Rev. Lett.}\ }\textbf
  {\bibinfo {volume} {90}},\ \bibinfo {pages} {074302} (\bibinfo {year}
  {2003})}\BibitemShut {NoStop}%
\bibitem [{\citenamefont {Audoly}\ and\ \citenamefont
  {Pomeau}(2010)}]{audoly2010from}%
  \BibitemOpen
  \bibfield  {author} {\bibinfo {author} {\bibfnamefont {B.}~\bibnamefont
  {Audoly}}\ and\ \bibinfo {author} {\bibfnamefont {Y.}~\bibnamefont
  {Pomeau}},\ }\href@noop {} {\emph {\bibinfo {title} {Elasticity and geometry:
  {F}rom hair curls to the non-linear response of shells}}}\ (\bibinfo
  {publisher} {Oxford University Press, Oxford},\ \bibinfo {year}
  {2010})\BibitemShut {NoStop}%
\bibitem [{\citenamefont {Shyer}\ \emph {et~al.}(2013)\citenamefont {Shyer},
  \citenamefont {Tallinen}, \citenamefont {Nerurkar}, \citenamefont {Wei},
  \citenamefont {Gil}, \citenamefont {Kaplan}, \citenamefont {Tabin},\ and\
  \citenamefont {Mahadevan}}]{shyer2013villification}%
  \BibitemOpen
  \bibfield  {author} {\bibinfo {author} {\bibfnamefont {A.~E.}\ \bibnamefont
  {Shyer}}, \bibinfo {author} {\bibfnamefont {T.}~\bibnamefont {Tallinen}},
  \bibinfo {author} {\bibfnamefont {N.~L.}\ \bibnamefont {Nerurkar}}, \bibinfo
  {author} {\bibfnamefont {Z.}~\bibnamefont {Wei}}, \bibinfo {author}
  {\bibfnamefont {E.~S.}\ \bibnamefont {Gil}}, \bibinfo {author} {\bibfnamefont
  {D.~L.}\ \bibnamefont {Kaplan}}, \bibinfo {author} {\bibfnamefont {C.~J.}\
  \bibnamefont {Tabin}}, \ and\ \bibinfo {author} {\bibfnamefont
  {L.}~\bibnamefont {Mahadevan}},\ }\bibfield  {title} {\enquote {\bibinfo
  {title} {Villification: {H}ow the gut gets its villi},}\ }\href {\doibase
  10.1126/science.1238842} {\bibfield  {journal} {\bibinfo  {journal}
  {Science}\ }\textbf {\bibinfo {volume} {342}},\ \bibinfo {pages} {212--218}
  (\bibinfo {year} {2013})}\BibitemShut {NoStop}%
\bibitem [{\citenamefont {Gemmer}\ \emph {et~al.}(2016)\citenamefont {Gemmer},
  \citenamefont {Sharon}, \citenamefont {Shearman},\ and\ \citenamefont
  {Venkataramani}}]{Gemmer2016}%
  \BibitemOpen
  \bibfield  {author} {\bibinfo {author} {\bibfnamefont {J.}~\bibnamefont
  {Gemmer}}, \bibinfo {author} {\bibfnamefont {E.}~\bibnamefont {Sharon}},
  \bibinfo {author} {\bibfnamefont {T.}~\bibnamefont {Shearman}}, \ and\
  \bibinfo {author} {\bibfnamefont {S.~C.}\ \bibnamefont {Venkataramani}},\
  }\bibfield  {title} {\enquote {\bibinfo {title} {Isometric immersions, energy
  minimization and self-similar buckling in non-{E}uclidean elastic sheets},}\
  }\href {\doibase 10.1209/0295-5075/114/24003} {\bibfield  {journal} {\bibinfo
   {journal} {EPL}\ }\textbf {\bibinfo {volume} {114}},\ \bibinfo {pages}
  {24003} (\bibinfo {year} {2016})}\BibitemShut {NoStop}%
\bibitem [{\citenamefont {Xu}\ \emph {et~al.}(2020)\citenamefont {Xu},
  \citenamefont {Fu},\ and\ \citenamefont {Yang}}]{Xu2020water}%
  \BibitemOpen
  \bibfield  {author} {\bibinfo {author} {\bibfnamefont {F.}~\bibnamefont
  {Xu}}, \bibinfo {author} {\bibfnamefont {C.}~\bibnamefont {Fu}}, \ and\
  \bibinfo {author} {\bibfnamefont {Y.}~\bibnamefont {Yang}},\ }\bibfield
  {title} {\enquote {\bibinfo {title} {Water affects morphogenesis of growing
  aquatic plant leaves},}\ }\href {\doibase 10.1103/PhysRevLett.124.038003}
  {\bibfield  {journal} {\bibinfo  {journal} {Phys. Rev. Lett.}\ }\textbf
  {\bibinfo {volume} {124}},\ \bibinfo {pages} {038003} (\bibinfo {year}
  {2020})}\BibitemShut {NoStop}%
\bibitem [{\citenamefont {Fei}\ \emph {et~al.}(2020)\citenamefont {Fei},
  \citenamefont {Mao}, \citenamefont {Yan}, \citenamefont {Alert},
  \citenamefont {Stone}, \citenamefont {Bassler}, \citenamefont {Wingreen},\
  and\ \citenamefont {Ko{\v{s}}mrlj}}]{Fei2020}%
  \BibitemOpen
  \bibfield  {author} {\bibinfo {author} {\bibfnamefont {C.}~\bibnamefont
  {Fei}}, \bibinfo {author} {\bibfnamefont {S.}~\bibnamefont {Mao}}, \bibinfo
  {author} {\bibfnamefont {J.}~\bibnamefont {Yan}}, \bibinfo {author}
  {\bibfnamefont {R.}~\bibnamefont {Alert}}, \bibinfo {author} {\bibfnamefont
  {H.~A.}\ \bibnamefont {Stone}}, \bibinfo {author} {\bibfnamefont {B.~L.}\
  \bibnamefont {Bassler}}, \bibinfo {author} {\bibfnamefont {N.~S.}\
  \bibnamefont {Wingreen}}, \ and\ \bibinfo {author} {\bibfnamefont
  {A.}~\bibnamefont {Ko{\v{s}}mrlj}},\ }\bibfield  {title} {\enquote {\bibinfo
  {title} {{Nonuniform growth and surface friction determine bacterial biofilm
  morphology on soft substrates}},}\ }\href {\doibase 10.1073/pnas.1919607117}
  {\bibfield  {journal} {\bibinfo  {journal} {Proc. Natl. Acad. Sci.}\ }\textbf
  {\bibinfo {volume} {117}},\ \bibinfo {pages} {7622--7632} (\bibinfo {year}
  {2020})}\BibitemShut {NoStop}%
\bibitem [{\citenamefont {Hure}\ \emph {et~al.}(2012)\citenamefont {Hure},
  \citenamefont {Roman},\ and\ \citenamefont {Bico}}]{Hure12}%
  \BibitemOpen
  \bibfield  {author} {\bibinfo {author} {\bibfnamefont {J.}~\bibnamefont
  {Hure}}, \bibinfo {author} {\bibfnamefont {B.}~\bibnamefont {Roman}}, \ and\
  \bibinfo {author} {\bibfnamefont {J.}~\bibnamefont {Bico}},\ }\bibfield
  {title} {\enquote {\bibinfo {title} {Stamping and wrinkling of elastic
  plates},}\ }\href {\doibase 10.1103/PhysRevLett.109.054302} {\bibfield
  {journal} {\bibinfo  {journal} {Phys. Rev. Lett.}\ }\textbf {\bibinfo
  {volume} {109}},\ \bibinfo {pages} {054302} (\bibinfo {year}
  {2012})}\BibitemShut {NoStop}%
\bibitem [{\citenamefont {King}\ \emph {et~al.}(2012)\citenamefont {King},
  \citenamefont {Schroll}, \citenamefont {Davidovitch},\ and\ \citenamefont
  {Menon}}]{King12}%
  \BibitemOpen
  \bibfield  {author} {\bibinfo {author} {\bibfnamefont {H.}~\bibnamefont
  {King}}, \bibinfo {author} {\bibfnamefont {R.~D.}\ \bibnamefont {Schroll}},
  \bibinfo {author} {\bibfnamefont {B.}~\bibnamefont {Davidovitch}}, \ and\
  \bibinfo {author} {\bibfnamefont {N.}~\bibnamefont {Menon}},\ }\bibfield
  {title} {\enquote {\bibinfo {title} {Elastic sheet on a liquid drop reveals
  wrinkling and crumpling as distinct symmetry-breaking instabilities},}\
  }\href {\doibase 10.1073/pnas.1201201109} {\bibfield  {journal} {\bibinfo
  {journal} {Proc. Natl. Acad. Sci.}\ }\textbf {\bibinfo {volume} {109}},\
  \bibinfo {pages} {9716--9720} (\bibinfo {year} {2012})}\BibitemShut {NoStop}%
\bibitem [{\citenamefont {Paulsen}(2019)}]{Paulsen19}%
  \BibitemOpen
  \bibfield  {author} {\bibinfo {author} {\bibfnamefont {J.~D.}\ \bibnamefont
  {Paulsen}},\ }\bibfield  {title} {\enquote {\bibinfo {title} {Wrapping
  liquids, solids, and gases in thin sheets},}\ }\href {\doibase
  10.1146/annurev-conmatphys-031218-013533} {\bibfield  {journal} {\bibinfo
  {journal} {Annu. Rev. Condens. Matter Phys.}\ }\textbf {\bibinfo {volume}
  {10}},\ \bibinfo {pages} {431--450} (\bibinfo {year} {2019})}\BibitemShut
  {NoStop}%
\bibitem [{\citenamefont {Vella}(2019)}]{Vella2019buffering}%
  \BibitemOpen
  \bibfield  {author} {\bibinfo {author} {\bibfnamefont {D.}~\bibnamefont
  {Vella}},\ }\bibfield  {title} {\enquote {\bibinfo {title} {Buffering by
  buckling as a route for elastic deformation},}\ }\href {\doibase
  10.1038/s42254-019-0063-1} {\bibfield  {journal} {\bibinfo  {journal} {Nat.
  Rev. Phys.}\ }\textbf {\bibinfo {volume} {1}},\ \bibinfo {pages} {425--436}
  (\bibinfo {year} {2019})}\BibitemShut {NoStop}%
\bibitem [{\citenamefont {Timounay}\ \emph {et~al.}(2021)\citenamefont
  {Timounay}, \citenamefont {Hartwell}, \citenamefont {He}, \citenamefont
  {King}, \citenamefont {Murphy}, \citenamefont {D\'emery},\ and\ \citenamefont
  {Paulsen}}]{Timounay21}%
  \BibitemOpen
  \bibfield  {author} {\bibinfo {author} {\bibfnamefont {Yousra}\ \bibnamefont
  {Timounay}}, \bibinfo {author} {\bibfnamefont {Alexander~R.}\ \bibnamefont
  {Hartwell}}, \bibinfo {author} {\bibfnamefont {Mengfei}\ \bibnamefont {He}},
  \bibinfo {author} {\bibfnamefont {D.~Eric}\ \bibnamefont {King}}, \bibinfo
  {author} {\bibfnamefont {Lindsay~K.}\ \bibnamefont {Murphy}}, \bibinfo
  {author} {\bibfnamefont {Vincent}\ \bibnamefont {D\'emery}}, \ and\ \bibinfo
  {author} {\bibfnamefont {Joseph~D.}\ \bibnamefont {Paulsen}},\ }\bibfield
  {title} {\enquote {\bibinfo {title} {Sculpting liquids with ultrathin
  shells},}\ }\href {\doibase 10.1103/PhysRevLett.127.108002} {\bibfield
  {journal} {\bibinfo  {journal} {Phys. Rev. Lett.}\ }\textbf {\bibinfo
  {volume} {127}},\ \bibinfo {pages} {108002} (\bibinfo {year}
  {2021})}\BibitemShut {NoStop}%
\bibitem [{\citenamefont {Breid}\ and\ \citenamefont {Crosby}(2013)}]{Breid13}%
  \BibitemOpen
  \bibfield  {author} {\bibinfo {author} {\bibfnamefont {D.}~\bibnamefont
  {Breid}}\ and\ \bibinfo {author} {\bibfnamefont {A.~J.}\ \bibnamefont
  {Crosby}},\ }\bibfield  {title} {\enquote {\bibinfo {title}
  {Curvature-controlled wrinkle morphologies},}\ }\href {\doibase
  10.1039/C3SM27331H} {\bibfield  {journal} {\bibinfo  {journal} {Soft Matter}\
  }\textbf {\bibinfo {volume} {9}},\ \bibinfo {pages} {3624--3630} (\bibinfo
  {year} {2013})}\BibitemShut {NoStop}%
\bibitem [{\citenamefont {Stoop}\ \emph {et~al.}(2015)\citenamefont {Stoop},
  \citenamefont {Lagrange}, \citenamefont {Terwagne}, \citenamefont {Reis},\
  and\ \citenamefont {Dunkel}}]{Stoop15}%
  \BibitemOpen
  \bibfield  {author} {\bibinfo {author} {\bibfnamefont {N.}~\bibnamefont
  {Stoop}}, \bibinfo {author} {\bibfnamefont {R.}~\bibnamefont {Lagrange}},
  \bibinfo {author} {\bibfnamefont {D.}~\bibnamefont {Terwagne}}, \bibinfo
  {author} {\bibfnamefont {P.~M.}\ \bibnamefont {Reis}}, \ and\ \bibinfo
  {author} {\bibfnamefont {J.}~\bibnamefont {Dunkel}},\ }\bibfield  {title}
  {\enquote {\bibinfo {title} {Curvature-induced symmetry breaking determines
  elastic surface patterns},}\ }\href {\doibase 10.1038/nmat4202} {\bibfield
  {journal} {\bibinfo  {journal} {Nat. Mater.}\ }\textbf {\bibinfo {volume}
  {14}},\ \bibinfo {pages} {337--342} (\bibinfo {year} {2015})}\BibitemShut
  {NoStop}%
\bibitem [{\citenamefont {Reis}(2015)}]{Reis2015}%
  \BibitemOpen
  \bibfield  {author} {\bibinfo {author} {\bibfnamefont {P.~M.}\ \bibnamefont
  {Reis}},\ }\bibfield  {title} {\enquote {\bibinfo {title} {A perspective on
  the revival of structural (in)stability with novel opportunities for
  function: {F}rom buckliphobia to buckliphilia},}\ }\href {\doibase
  10.1115/1.4031456} {\bibfield  {journal} {\bibinfo  {journal} {J. Appl.
  Mech.}\ }\textbf {\bibinfo {volume} {82}},\ \bibinfo {pages} {111001}
  (\bibinfo {year} {2015})}\BibitemShut {NoStop}%
\bibitem [{\citenamefont {Aharoni}\ \emph {et~al.}(2017)\citenamefont
  {Aharoni}, \citenamefont {Todorova}, \citenamefont {Albarr{\'a}n},
  \citenamefont {Goehring}, \citenamefont {Kamien},\ and\ \citenamefont
  {Katifori}}]{Aharoni17}%
  \BibitemOpen
  \bibfield  {author} {\bibinfo {author} {\bibfnamefont {H.}~\bibnamefont
  {Aharoni}}, \bibinfo {author} {\bibfnamefont {D.~V.}\ \bibnamefont
  {Todorova}}, \bibinfo {author} {\bibfnamefont {O.}~\bibnamefont
  {Albarr{\'a}n}}, \bibinfo {author} {\bibfnamefont {L.}~\bibnamefont
  {Goehring}}, \bibinfo {author} {\bibfnamefont {R.~D.}\ \bibnamefont
  {Kamien}}, \ and\ \bibinfo {author} {\bibfnamefont {E.}~\bibnamefont
  {Katifori}},\ }\bibfield  {title} {\enquote {\bibinfo {title} {The smectic
  order of wrinkles},}\ }\href {\doibase 10.1038/ncomms15809} {\bibfield
  {journal} {\bibinfo  {journal} {Nat. Commun.}\ }\textbf {\bibinfo {volume}
  {8}},\ \bibinfo {pages} {15809} (\bibinfo {year} {2017})}\BibitemShut
  {NoStop}%
\bibitem [{\citenamefont {Bella}\ and\ \citenamefont
  {Kohn}(2017)}]{bella2017wrinkling}%
  \BibitemOpen
  \bibfield  {author} {\bibinfo {author} {\bibfnamefont {P.}~\bibnamefont
  {Bella}}\ and\ \bibinfo {author} {\bibfnamefont {R.~V.}\ \bibnamefont
  {Kohn}},\ }\bibfield  {title} {\enquote {\bibinfo {title} {Wrinkling of a
  thin circular sheet bonded to a spherical substrate},}\ }\href {\doibase
  10.1098/rsta.2016.0157} {\bibfield  {journal} {\bibinfo  {journal} {Philos.
  Trans. R. Soc. A}\ }\textbf {\bibinfo {volume} {375}},\ \bibinfo {pages}
  {20160157} (\bibinfo {year} {2017})}\BibitemShut {NoStop}%
\bibitem [{\citenamefont {Zhang}\ \emph {et~al.}(2019)\citenamefont {Zhang},
  \citenamefont {Mather}, \citenamefont {Bowick},\ and\ \citenamefont
  {Zhang}}]{zhang2019nonuniform}%
  \BibitemOpen
  \bibfield  {author} {\bibinfo {author} {\bibfnamefont {X.}~\bibnamefont
  {Zhang}}, \bibinfo {author} {\bibfnamefont {P.~T.}\ \bibnamefont {Mather}},
  \bibinfo {author} {\bibfnamefont {M.~J.}\ \bibnamefont {Bowick}}, \ and\
  \bibinfo {author} {\bibfnamefont {T.}~\bibnamefont {Zhang}},\ }\bibfield
  {title} {\enquote {\bibinfo {title} {Non-uniform curvature and anisotropic
  deformation control wrinkling patterns on tori},}\ }\href {\doibase
  10.1039/C9SM00235A} {\bibfield  {journal} {\bibinfo  {journal} {Soft Matter}\
  }\textbf {\bibinfo {volume} {15}},\ \bibinfo {pages} {5204--5210} (\bibinfo
  {year} {2019})}\BibitemShut {NoStop}%
\bibitem [{\citenamefont {Davidovitch}\ \emph {et~al.}(2019)\citenamefont
  {Davidovitch}, \citenamefont {Sun},\ and\ \citenamefont
  {Grason}}]{davidovitch2019}%
  \BibitemOpen
  \bibfield  {author} {\bibinfo {author} {\bibfnamefont {B.}~\bibnamefont
  {Davidovitch}}, \bibinfo {author} {\bibfnamefont {Y.}~\bibnamefont {Sun}}, \
  and\ \bibinfo {author} {\bibfnamefont {G.~M.}\ \bibnamefont {Grason}},\
  }\bibfield  {title} {\enquote {\bibinfo {title} {Geometrically incompatible
  confinement of solids},}\ }\href@noop {} {\bibfield  {journal} {\bibinfo
  {journal} {Proceedings of the National Academy of Sciences}\ }\textbf
  {\bibinfo {volume} {116}},\ \bibinfo {pages} {1483--1488} (\bibinfo {year}
  {2019})}\BibitemShut {NoStop}%
\bibitem [{\citenamefont {Tovkach}\ \emph {et~al.}(2020)\citenamefont
  {Tovkach}, \citenamefont {Chen}, \citenamefont {Ripp}, \citenamefont {Zhang},
  \citenamefont {Paulsen},\ and\ \citenamefont {Davidovitch}}]{Tovkach2020}%
  \BibitemOpen
  \bibfield  {author} {\bibinfo {author} {\bibfnamefont {O.}~\bibnamefont
  {Tovkach}}, \bibinfo {author} {\bibfnamefont {J.}~\bibnamefont {Chen}},
  \bibinfo {author} {\bibfnamefont {M.~M.}\ \bibnamefont {Ripp}}, \bibinfo
  {author} {\bibfnamefont {T.}~\bibnamefont {Zhang}}, \bibinfo {author}
  {\bibfnamefont {J.~D.}\ \bibnamefont {Paulsen}}, \ and\ \bibinfo {author}
  {\bibfnamefont {B.}~\bibnamefont {Davidovitch}},\ }\bibfield  {title}
  {\enquote {\bibinfo {title} {Mesoscale structure of wrinkle patterns and
  defect-proliferated liquid crystalline phases},}\ }\href {\doibase
  10.1073/pnas.1916221117} {\bibfield  {journal} {\bibinfo  {journal} {Proc.
  Natl. Acad. Sci.}\ }\textbf {\bibinfo {volume} {117}},\ \bibinfo {pages}
  {3938--3943} (\bibinfo {year} {2020})}\BibitemShut {NoStop}%
\bibitem [{\citenamefont {Pretzl}\ \emph {et~al.}(2008)\citenamefont {Pretzl},
  \citenamefont {Schweikart}, \citenamefont {Hanske}, \citenamefont {Chiche},
  \citenamefont {Zettl}, \citenamefont {Horn}, \citenamefont {B\"oker},\ and\
  \citenamefont {Fery}}]{pretzl2008lithography}%
  \BibitemOpen
  \bibfield  {author} {\bibinfo {author} {\bibfnamefont {M.}~\bibnamefont
  {Pretzl}}, \bibinfo {author} {\bibfnamefont {A.}~\bibnamefont {Schweikart}},
  \bibinfo {author} {\bibfnamefont {C.}~\bibnamefont {Hanske}}, \bibinfo
  {author} {\bibfnamefont {A.}~\bibnamefont {Chiche}}, \bibinfo {author}
  {\bibfnamefont {U.}~\bibnamefont {Zettl}}, \bibinfo {author} {\bibfnamefont
  {A.}~\bibnamefont {Horn}}, \bibinfo {author} {\bibfnamefont {A.}~\bibnamefont
  {B\"oker}}, \ and\ \bibinfo {author} {\bibfnamefont {A.}~\bibnamefont
  {Fery}},\ }\bibfield  {title} {\enquote {\bibinfo {title} {A lithography-free
  pathway for chemical microstructuring of macromolecules from aqueous solution
  based on wrinkling},}\ }\href {\doibase 10.1021/la8021694} {\bibfield
  {journal} {\bibinfo  {journal} {Langmuir}\ }\textbf {\bibinfo {volume}
  {24}},\ \bibinfo {pages} {12748--12753} (\bibinfo {year} {2008})}\BibitemShut
  {NoStop}%
\bibitem [{\citenamefont {Yang}\ \emph {et~al.}(2010)\citenamefont {Yang},
  \citenamefont {Khare},\ and\ \citenamefont {Lin}}]{Yang2010}%
  \BibitemOpen
  \bibfield  {author} {\bibinfo {author} {\bibfnamefont {S.}~\bibnamefont
  {Yang}}, \bibinfo {author} {\bibfnamefont {K.}~\bibnamefont {Khare}}, \ and\
  \bibinfo {author} {\bibfnamefont {P.-C.}\ \bibnamefont {Lin}},\ }\bibfield
  {title} {\enquote {\bibinfo {title} {{Harnessing Surface Wrinkle Patterns in
  Soft Matter}},}\ }\href {\doibase 10.1002/adfm.201000034} {\bibfield
  {journal} {\bibinfo  {journal} {Adv. Funct. Mater.}\ }\textbf {\bibinfo
  {volume} {20}},\ \bibinfo {pages} {2550--2564} (\bibinfo {year}
  {2010})}\BibitemShut {NoStop}%
\bibitem [{\citenamefont {Chen}\ and\ \citenamefont {Yang}(2012)}]{Chen2012}%
  \BibitemOpen
  \bibfield  {author} {\bibinfo {author} {\bibfnamefont {C.-M.}\ \bibnamefont
  {Chen}}\ and\ \bibinfo {author} {\bibfnamefont {S.}~\bibnamefont {Yang}},\
  }\bibfield  {title} {\enquote {\bibinfo {title} {{Wrinkling instabilities in
  polymer films and their applications}},}\ }\href {\doibase 10.1002/pi.4223}
  {\bibfield  {journal} {\bibinfo  {journal} {Polym. Int.}\ }\textbf {\bibinfo
  {volume} {61}},\ \bibinfo {pages} {1041--1047} (\bibinfo {year}
  {2012})}\BibitemShut {NoStop}%
\bibitem [{\citenamefont {Li}\ \emph {et~al.}(2017)\citenamefont {Li},
  \citenamefont {Zhai}, \citenamefont {Wang}, \citenamefont {Wendland},
  \citenamefont {Yin},\ and\ \citenamefont {Xiao}}]{Li2017}%
  \BibitemOpen
  \bibfield  {author} {\bibinfo {author} {\bibfnamefont {Z.}~\bibnamefont
  {Li}}, \bibinfo {author} {\bibfnamefont {Y.}~\bibnamefont {Zhai}}, \bibinfo
  {author} {\bibfnamefont {Y.}~\bibnamefont {Wang}}, \bibinfo {author}
  {\bibfnamefont {G.~M.}\ \bibnamefont {Wendland}}, \bibinfo {author}
  {\bibfnamefont {X.}~\bibnamefont {Yin}}, \ and\ \bibinfo {author}
  {\bibfnamefont {J.}~\bibnamefont {Xiao}},\ }\bibfield  {title} {\enquote
  {\bibinfo {title} {Harnessing surface wrinkling--cracking patterns for
  tunable optical transmittance},}\ }\href {\doibase 10.1002/adom.201700425}
  {\bibfield  {journal} {\bibinfo  {journal} {Adv. Opt. Mater.}\ }\textbf
  {\bibinfo {volume} {5}},\ \bibinfo {pages} {1--7} (\bibinfo {year}
  {2017})}\BibitemShut {NoStop}%
\bibitem [{\citenamefont {Wagner}(1929)}]{wagner1929ebene}%
  \BibitemOpen
  \bibfield  {author} {\bibinfo {author} {\bibfnamefont {H.}~\bibnamefont
  {Wagner}},\ }\bibfield  {title} {\enquote {\bibinfo {title} {Ebene
  blechwandtr\"{a}ger mit sehr d\"{u}nnem stegblech},}\ }\href@noop {}
  {\bibfield  {journal} {\bibinfo  {journal} {Z. Flugtech. Motorluftshiffahrt}\
  }\textbf {\bibinfo {volume} {20}},\ \bibinfo {pages} {200} (\bibinfo {year}
  {1929})}\BibitemShut {NoStop}%
\bibitem [{\citenamefont {Pipkin}(1986)}]{Pipkin86}%
  \BibitemOpen
  \bibfield  {author} {\bibinfo {author} {\bibfnamefont {A.C.}\ \bibnamefont
  {Pipkin}},\ }\bibfield  {title} {\enquote {\bibinfo {title} {The relaxed
  energy density for isotropic elastic membranes},}\ }\href {\doibase
  10.1093/imamat/36.1.85} {\bibfield  {journal} {\bibinfo  {journal} {IMA J.
  Appl. Math.}\ }\textbf {\bibinfo {volume} {36}},\ \bibinfo {pages} {85--99}
  (\bibinfo {year} {1986})}\BibitemShut {NoStop}%
\bibitem [{\citenamefont {Steigmann}(1990)}]{Steigmann90}%
  \BibitemOpen
  \bibfield  {author} {\bibinfo {author} {\bibfnamefont {D.~J.}\ \bibnamefont
  {Steigmann}},\ }\bibfield  {title} {\enquote {\bibinfo {title} {Tension-field
  theory},}\ }\href {\doibase 10.1098/rspa.1990.0055} {\bibfield  {journal}
  {\bibinfo  {journal} {Proc. Roy. Soc. London Ser. A}\ }\textbf {\bibinfo
  {volume} {429}},\ \bibinfo {pages} {141--173} (\bibinfo {year}
  {1990})}\BibitemShut {NoStop}%
\bibitem [{\citenamefont {Davidovitch}\ \emph {et~al.}(2011)\citenamefont
  {Davidovitch}, \citenamefont {Schroll}, \citenamefont {Vella}, \citenamefont
  {Adda-Bedia},\ and\ \citenamefont {Cerda}}]{Davidovitch11}%
  \BibitemOpen
  \bibfield  {author} {\bibinfo {author} {\bibfnamefont {Benny}\ \bibnamefont
  {Davidovitch}}, \bibinfo {author} {\bibfnamefont {Robert~D.}\ \bibnamefont
  {Schroll}}, \bibinfo {author} {\bibfnamefont {Dominic}\ \bibnamefont
  {Vella}}, \bibinfo {author} {\bibfnamefont {Mokhtar}\ \bibnamefont
  {Adda-Bedia}}, \ and\ \bibinfo {author} {\bibfnamefont {Enrique~A.}\
  \bibnamefont {Cerda}},\ }\bibfield  {title} {\enquote {\bibinfo {title}
  {Prototypical model for tensional wrinkling in thin sheets},}\ }\href
  {\doibase 10.1073/pnas.1108553108} {\bibfield  {journal} {\bibinfo  {journal}
  {Proc.\ Natl.\ Acad.\ Sci.}\ }\textbf {\bibinfo {volume} {108}},\ \bibinfo
  {pages} {18227--18232} (\bibinfo {year} {2011})}\BibitemShut {NoStop}%
\bibitem [{\citenamefont {Bella}\ and\ \citenamefont
  {Kohn}(2014)}]{BellaKohn2014}%
  \BibitemOpen
  \bibfield  {author} {\bibinfo {author} {\bibfnamefont {P.}~\bibnamefont
  {Bella}}\ and\ \bibinfo {author} {\bibfnamefont {R.~V.}\ \bibnamefont
  {Kohn}},\ }\bibfield  {title} {\enquote {\bibinfo {title} {Wrinkles as the
  result of compressive stresses in an annular thin film},}\ }\href {\doibase
  10.1002/cpa.21471} {\bibfield  {journal} {\bibinfo  {journal} {Comm. Pure
  Appl. Math.}\ }\textbf {\bibinfo {volume} {67}},\ \bibinfo {pages} {693--747}
  (\bibinfo {year} {2014})}\BibitemShut {NoStop}%
\bibitem [{\citenamefont {Hohlfeld}\ and\ \citenamefont
  {Davidovitch}(2015)}]{Hohlfeld15}%
  \BibitemOpen
  \bibfield  {author} {\bibinfo {author} {\bibfnamefont {E.}~\bibnamefont
  {Hohlfeld}}\ and\ \bibinfo {author} {\bibfnamefont {B.}~\bibnamefont
  {Davidovitch}},\ }\bibfield  {title} {\enquote {\bibinfo {title} {Sheet on a
  deformable sphere: Wrinkle patterns suppress curvature-induced
  delamination},}\ }\href {\doibase 10.1103/PhysRevE.91.012407} {\bibfield
  {journal} {\bibinfo  {journal} {Phys. Rev. E}\ }\textbf {\bibinfo {volume}
  {91}},\ \bibinfo {pages} {012407} (\bibinfo {year} {2015})}\BibitemShut
  {NoStop}%
\bibitem [{\citenamefont {Vella}\ \emph {et~al.}(2015)\citenamefont {Vella},
  \citenamefont {Huang}, \citenamefont {Menon}, \citenamefont {Russell},\ and\
  \citenamefont {Davidovitch}}]{Vella15}%
  \BibitemOpen
  \bibfield  {author} {\bibinfo {author} {\bibfnamefont {D.}~\bibnamefont
  {Vella}}, \bibinfo {author} {\bibfnamefont {J.}~\bibnamefont {Huang}},
  \bibinfo {author} {\bibfnamefont {N.}~\bibnamefont {Menon}}, \bibinfo
  {author} {\bibfnamefont {T.~P.}\ \bibnamefont {Russell}}, \ and\ \bibinfo
  {author} {\bibfnamefont {B.}~\bibnamefont {Davidovitch}},\ }\bibfield
  {title} {\enquote {\bibinfo {title} {Indentation of ultrathin elastic films
  and the emergence of asymptotic isometry},}\ }\href {\doibase
  10.1103/PhysRevLett.114.014301} {\bibfield  {journal} {\bibinfo  {journal}
  {Phys. Rev. Lett.}\ }\textbf {\bibinfo {volume} {114}},\ \bibinfo {pages}
  {014301} (\bibinfo {year} {2015})}\BibitemShut {NoStop}%
\bibitem [{\citenamefont {Taffetani}\ and\ \citenamefont
  {Vella}(2017)}]{Taffetani17}%
  \BibitemOpen
  \bibfield  {author} {\bibinfo {author} {\bibfnamefont {M.}~\bibnamefont
  {Taffetani}}\ and\ \bibinfo {author} {\bibfnamefont {D.}~\bibnamefont
  {Vella}},\ }\bibfield  {title} {\enquote {\bibinfo {title} {Regimes of
  wrinkling in pressurized elastic shells},}\ }\href {\doibase
  10.1098/rsta.2016.0330} {\bibfield  {journal} {\bibinfo  {journal} {Philos.
  Trans. Roy. Soc. A}\ }\textbf {\bibinfo {volume} {375}},\ \bibinfo {pages}
  {20160330} (\bibinfo {year} {2017})}\BibitemShut {NoStop}%
\bibitem [{\citenamefont {Amar}\ and\ \citenamefont {Jia}(2013)}]{Amar13}%
  \BibitemOpen
  \bibfield  {author} {\bibinfo {author} {\bibfnamefont {Martine~Ben}\
  \bibnamefont {Amar}}\ and\ \bibinfo {author} {\bibfnamefont {Fei}\
  \bibnamefont {Jia}},\ }\bibfield  {title} {\enquote {\bibinfo {title}
  {Anisotropic growth shapes intestinal tissues during embryogenesis},}\
  }\href@noop {} {\bibfield  {journal} {\bibinfo  {journal} {Proceedings of the
  National Academy of Sciences}\ }\textbf {\bibinfo {volume} {110}},\ \bibinfo
  {pages} {10525--10530} (\bibinfo {year} {2013})}\BibitemShut {NoStop}%
\bibitem [{\citenamefont {van Rees}\ \emph {et~al.}(2017)\citenamefont {van
  Rees}, \citenamefont {Vouga},\ and\ \citenamefont {Mahadevan}}]{vanRees17}%
  \BibitemOpen
  \bibfield  {author} {\bibinfo {author} {\bibfnamefont {Wim~M}\ \bibnamefont
  {van Rees}}, \bibinfo {author} {\bibfnamefont {Etienne}\ \bibnamefont
  {Vouga}}, \ and\ \bibinfo {author} {\bibfnamefont {Lakshminarayanan}\
  \bibnamefont {Mahadevan}},\ }\bibfield  {title} {\enquote {\bibinfo {title}
  {Growth patterns for shape-shifting elastic bilayers},}\ }\href@noop {}
  {\bibfield  {journal} {\bibinfo  {journal} {Proceedings of the National
  Academy of Sciences}\ }\textbf {\bibinfo {volume} {114}},\ \bibinfo {pages}
  {11597--11602} (\bibinfo {year} {2017})}\BibitemShut {NoStop}%
\bibitem [{\citenamefont {Tobasco}(2021)}]{tobasco2021curvature}%
  \BibitemOpen
  \bibfield  {author} {\bibinfo {author} {\bibfnamefont {I.}~\bibnamefont
  {Tobasco}},\ }\bibfield  {title} {\enquote {\bibinfo {title}
  {Curvature-driven wrinkling of thin elastic shells},}\ }\href {\doibase
  10.1007/s00205-020-01566-8} {\bibfield  {journal} {\bibinfo  {journal} {Arch.
  Ration. Mech. Anal.}\ }\textbf {\bibinfo {volume} {239}},\ \bibinfo {pages}
  {1211--1325} (\bibinfo {year} {2021})}\BibitemShut {NoStop}%
\bibitem [{\citenamefont {Ciarlet}(1997)}]{ciarlet1997mathematical}%
  \BibitemOpen
  \bibfield  {author} {\bibinfo {author} {\bibfnamefont {Philippe~G.}\
  \bibnamefont {Ciarlet}},\ }\href@noop {} {\emph {\bibinfo {title}
  {Mathematical elasticity. {V}ol. {II}}}},\ \bibinfo {series} {Studies in
  Mathematics and its Applications}, Vol.~\bibinfo {volume} {27}\ (\bibinfo
  {publisher} {North-Holland Publishing Co., Amsterdam},\ \bibinfo {year}
  {1997})\ pp.\ \bibinfo {pages} {lxiv+497},\ \bibinfo {note} {theory of
  plates}\BibitemShut {NoStop}%
\bibitem [{\citenamefont {Maxwell}(1864)}]{maxwell1864reciprocal}%
  \BibitemOpen
  \bibfield  {author} {\bibinfo {author} {\bibfnamefont {J.~C.}\ \bibnamefont
  {Maxwell}},\ }\bibfield  {title} {\enquote {\bibinfo {title} {{XLV. O}n
  reciprocal figures and diagrams of forces},}\ }\href {\doibase
  10.1080/14786446408643663} {\bibfield  {journal} {\bibinfo  {journal}
  {Philos. Mag.}\ }\textbf {\bibinfo {volume} {27}},\ \bibinfo {pages}
  {250--261} (\bibinfo {year} {1864})}\BibitemShut {NoStop}%
\bibitem [{\citenamefont {Timoshenko}(1953)}]{timoshenko1953history}%
  \BibitemOpen
  \bibfield  {author} {\bibinfo {author} {\bibfnamefont {S.~P.}\ \bibnamefont
  {Timoshenko}},\ }\href@noop {} {\emph {\bibinfo {title} {History of strength
  of materials. {W}ith a brief account of the history of theory of elasticity
  and theory of structures}}}\ (\bibinfo  {publisher} {McGraw-Hill Book
  Company, Inc., New York-Toronto-London},\ \bibinfo {year} {1953})\BibitemShut
  {NoStop}%
\bibitem [{\citenamefont {Prager}(1957)}]{Prager1957}%
  \BibitemOpen
  \bibfield  {author} {\bibinfo {author} {\bibfnamefont {W.}~\bibnamefont
  {Prager}},\ }\bibfield  {title} {\enquote {\bibinfo {title} {On ideal locking
  materials},}\ }\href {\doibase 10.1122/1.548818} {\bibfield  {journal}
  {\bibinfo  {journal} {Trans. Soc. Rheol.}\ }\textbf {\bibinfo {volume} {1}},\
  \bibinfo {pages} {169--175} (\bibinfo {year} {1957})}\BibitemShut {NoStop}%
\bibitem [{\citenamefont {Fung}(1967)}]{fung1967elasticity}%
  \BibitemOpen
  \bibfield  {author} {\bibinfo {author} {\bibfnamefont {Y.~C.}\ \bibnamefont
  {Fung}},\ }\bibfield  {title} {\enquote {\bibinfo {title} {Elasticity of soft
  tissues in simple elongation},}\ }\href {\doibase
  10.1152/ajplegacy.1967.213.6.1532} {\bibfield  {journal} {\bibinfo  {journal}
  {Am. J. Physiol.}\ }\textbf {\bibinfo {volume} {213}},\ \bibinfo {pages}
  {1532--1544} (\bibinfo {year} {1967})}\BibitemShut {NoStop}%
\bibitem [{\citenamefont {Prager}(1969)}]{Prager1969}%
  \BibitemOpen
  \bibfield  {author} {\bibinfo {author} {\bibfnamefont {W.}~\bibnamefont
  {Prager}},\ }\bibfield  {title} {\enquote {\bibinfo {title} {On the
  formulation of constitutive equations for living soft tissues},}\ }\href
  {\doibase 10.1090/qam/99834} {\bibfield  {journal} {\bibinfo  {journal} {Q.
  Appl. Math.}\ }\textbf {\bibinfo {volume} {27}},\ \bibinfo {pages} {128--132}
  (\bibinfo {year} {1969})}\BibitemShut {NoStop}%
\bibitem [{\citenamefont {Elettro}\ \emph {et~al.}(2016)\citenamefont
  {Elettro}, \citenamefont {Neukirch}, \citenamefont {Vollrath},\ and\
  \citenamefont {Antkowiak}}]{elettro2016drop}%
  \BibitemOpen
  \bibfield  {author} {\bibinfo {author} {\bibfnamefont {H.}~\bibnamefont
  {Elettro}}, \bibinfo {author} {\bibfnamefont {S.}~\bibnamefont {Neukirch}},
  \bibinfo {author} {\bibfnamefont {F.}~\bibnamefont {Vollrath}}, \ and\
  \bibinfo {author} {\bibfnamefont {A.}~\bibnamefont {Antkowiak}},\ }\bibfield
  {title} {\enquote {\bibinfo {title} {In-drop capillary spooling of spider
  capture thread inspires hybrid fibers with mixed solid--liquid mechanical
  properties},}\ }\href {\doibase 10.1073/pnas.1602451113} {\bibfield
  {journal} {\bibinfo  {journal} {Proc. Natl. Acad. Sci.}\ }\textbf {\bibinfo
  {volume} {113}},\ \bibinfo {pages} {6143--6147} (\bibinfo {year}
  {2016})}\BibitemShut {NoStop}%
\bibitem [{\citenamefont {Grandgeorge}\ \emph {et~al.}(2018)\citenamefont
  {Grandgeorge}, \citenamefont {Krins}, \citenamefont {Hourlier-Fargette},
  \citenamefont {Laberty-Robert}, \citenamefont {Neukirch},\ and\ \citenamefont
  {Antkowiak}}]{grandgeorge2018capillarity}%
  \BibitemOpen
  \bibfield  {author} {\bibinfo {author} {\bibfnamefont {P.}~\bibnamefont
  {Grandgeorge}}, \bibinfo {author} {\bibfnamefont {N.}~\bibnamefont {Krins}},
  \bibinfo {author} {\bibfnamefont {A.}~\bibnamefont {Hourlier-Fargette}},
  \bibinfo {author} {\bibfnamefont {C.}~\bibnamefont {Laberty-Robert}},
  \bibinfo {author} {\bibfnamefont {S.}~\bibnamefont {Neukirch}}, \ and\
  \bibinfo {author} {\bibfnamefont {A.}~\bibnamefont {Antkowiak}},\ }\bibfield
  {title} {\enquote {\bibinfo {title} {Capillarity-induced folds fuel extreme
  shape changes in thin wicked membranes},}\ }\href {\doibase
  10.1126/science.aaq0677} {\bibfield  {journal} {\bibinfo  {journal}
  {Science}\ }\textbf {\bibinfo {volume} {360}},\ \bibinfo {pages} {296--299}
  (\bibinfo {year} {2018})}\BibitemShut {NoStop}%
\bibitem [{\citenamefont {Timounay}\ \emph {et~al.}(2020)\citenamefont
  {Timounay}, \citenamefont {De}, \citenamefont {Stelzel}, \citenamefont
  {Schrecengost}, \citenamefont {Ripp},\ and\ \citenamefont
  {Paulsen}}]{Timounay20}%
  \BibitemOpen
  \bibfield  {author} {\bibinfo {author} {\bibfnamefont {Yousra}\ \bibnamefont
  {Timounay}}, \bibinfo {author} {\bibfnamefont {Raj}\ \bibnamefont {De}},
  \bibinfo {author} {\bibfnamefont {Jessica~L.}\ \bibnamefont {Stelzel}},
  \bibinfo {author} {\bibfnamefont {Zachariah~S.}\ \bibnamefont
  {Schrecengost}}, \bibinfo {author} {\bibfnamefont {Monica~M.}\ \bibnamefont
  {Ripp}}, \ and\ \bibinfo {author} {\bibfnamefont {Joseph~D.}\ \bibnamefont
  {Paulsen}},\ }\bibfield  {title} {\enquote {\bibinfo {title} {Crumples as a
  generic stress-focusing instability in confined sheets},}\ }\href {\doibase
  10.1103/PhysRevX.10.021008} {\bibfield  {journal} {\bibinfo  {journal} {Phys.
  Rev. X}\ }\textbf {\bibinfo {volume} {10}},\ \bibinfo {pages} {021008}
  (\bibinfo {year} {2020})}\BibitemShut {NoStop}%
\bibitem [{\citenamefont {Pocivavsek}\ \emph {et~al.}(2008)\citenamefont
  {Pocivavsek}, \citenamefont {Dellsy}, \citenamefont {Kern}, \citenamefont
  {Johnson}, \citenamefont {Lin}, \citenamefont {Lee},\ and\ \citenamefont
  {Cerda}}]{Pocivavsek08}%
  \BibitemOpen
  \bibfield  {author} {\bibinfo {author} {\bibfnamefont {Luka}\ \bibnamefont
  {Pocivavsek}}, \bibinfo {author} {\bibfnamefont {Robert}\ \bibnamefont
  {Dellsy}}, \bibinfo {author} {\bibfnamefont {Andrew}\ \bibnamefont {Kern}},
  \bibinfo {author} {\bibfnamefont {Sebasti{\'a}n}\ \bibnamefont {Johnson}},
  \bibinfo {author} {\bibfnamefont {Binhua}\ \bibnamefont {Lin}}, \bibinfo
  {author} {\bibfnamefont {Ka~Yee~C.}\ \bibnamefont {Lee}}, \ and\ \bibinfo
  {author} {\bibfnamefont {Enrique}\ \bibnamefont {Cerda}},\ }\bibfield
  {title} {\enquote {\bibinfo {title} {Stress and fold localization in thin
  elastic membranes},}\ }\href {\doibase 10.1126/science.1154069} {\bibfield
  {journal} {\bibinfo  {journal} {Science}\ }\textbf {\bibinfo {volume}
  {320}},\ \bibinfo {pages} {912--916} (\bibinfo {year} {2008})}\BibitemShut
  {NoStop}%
\bibitem [{\citenamefont {Brau}\ \emph {et~al.}(2013)\citenamefont {Brau},
  \citenamefont {Damman}, \citenamefont {Diamant},\ and\ \citenamefont
  {Witten}}]{Brau13}%
  \BibitemOpen
  \bibfield  {author} {\bibinfo {author} {\bibfnamefont {Fabian}\ \bibnamefont
  {Brau}}, \bibinfo {author} {\bibfnamefont {Pascal}\ \bibnamefont {Damman}},
  \bibinfo {author} {\bibfnamefont {Haim}\ \bibnamefont {Diamant}}, \ and\
  \bibinfo {author} {\bibfnamefont {Thomas~A.}\ \bibnamefont {Witten}},\
  }\bibfield  {title} {\enquote {\bibinfo {title} {Wrinkle to fold transition:
  influence of the substrate response},}\ }\href {\doibase 10.1039/C3SM50655J}
  {\bibfield  {journal} {\bibinfo  {journal} {Soft Matter}\ }\textbf {\bibinfo
  {volume} {9}},\ \bibinfo {pages} {8177--8186} (\bibinfo {year}
  {2013})}\BibitemShut {NoStop}%
\bibitem [{\citenamefont {Paulsen}\ \emph {et~al.}(2017)\citenamefont
  {Paulsen}, \citenamefont {D{\'e}mery}, \citenamefont {Toga}, \citenamefont
  {Qiu}, \citenamefont {Russell}, \citenamefont {Davidovitch},\ and\
  \citenamefont {Menon}}]{Paulsen17}%
  \BibitemOpen
  \bibfield  {author} {\bibinfo {author} {\bibfnamefont {J.~D.}\ \bibnamefont
  {Paulsen}}, \bibinfo {author} {\bibfnamefont {V.}~\bibnamefont {D{\'e}mery}},
  \bibinfo {author} {\bibfnamefont {K.~B.}\ \bibnamefont {Toga}}, \bibinfo
  {author} {\bibfnamefont {Z.}~\bibnamefont {Qiu}}, \bibinfo {author}
  {\bibfnamefont {T.~P.}\ \bibnamefont {Russell}}, \bibinfo {author}
  {\bibfnamefont {B.}~\bibnamefont {Davidovitch}}, \ and\ \bibinfo {author}
  {\bibfnamefont {N.}~\bibnamefont {Menon}},\ }\bibfield  {title} {\enquote
  {\bibinfo {title} {Geometry-driven folding of a floating annular sheet},}\
  }\href {\doibase 10.1103/PhysRevLett.118.048004} {\bibfield  {journal}
  {\bibinfo  {journal} {Phys. Rev. Lett.}\ }\textbf {\bibinfo {volume} {118}},\
  \bibinfo {pages} {048004} (\bibinfo {year} {2017})}\BibitemShut {NoStop}%
\end{thebibliography}%

\end{document}


 \global\long\def\thefigure{S\arabic{figure}}
\setcounter{figure}{0} 
\global\long\def\thetable{S\arabic{table}}
 \setcounter{table}{0} 
 \global\long\def\theequation{S\arabic{equation}}
 \setcounter{equation}{0}
 \global\long\def\thepage{S\arabic{page}}
 \setcounter{page}{1}

\title{Supplementary Information for: \\
 ``Exact solutions for the wrinkle patterns of confined elastic shells''}

\author{Tobasco et al.}

 \maketitle

 \tableofcontents




\section{Supplementary Tables}\label{sec:SITables}

\begin{table}[H]
\centering{}\begin{tabular}{l l c c c }
\hline
\hline
  & Planform shape \	&  $|R_i| \text{\,[mm]} $  	&  $t \text{\,[nm]}$   	&  $W \text{\,[mm]}$  		\\ 
\hline

\textbf{Fig.\ 1a} \ \  & Square 			& $25, 34$  				& 408			& 5.1				\\
\hline

\textbf{Fig.\ 2} \ \ 
& Ellipse	 		& $25, 34$				& 325			& 8.2				\\
& Tangential polygon  \	& $25, 34$				& 378			& 5.9				\\
\hline

\textbf{Fig.\ 3} \ \ & Triangle 		& $25, 34$				& 331			& 4.5				\\
& Rectangle 		& $25, 34$				& 242			& 5.0				\\
& Hexagon 		& $25, 34$				& 334			& 5.0				\\
& Tangential polygon  \	& $25, 34$				& 378			& 5.9				\\
& Semicircle 		& $25, 34$				& 389			& 6.3				\\
& Circle 			& $25, 34$				& 427			& 6.3				\\
& Ellipse	 		& $25, 34$				& 325			& 8.2				\\
\hline
\hline
\end{tabular}
\vspace{1em}
\caption{
Experimental parameters for the  saddle shells ($\kappa < 0$) in the main text. 
Unsigned principal radii of curvature $|R_1|$ and $|R_2|$, thickness $t$, and initial planar width $W$ are reported. 
}
\label{tab:exp-saddle}
\end{table}

\begin{table}[H]
\centering{}\begin{tabular}{l l c c c}
\hline
\hline
 & Planform shape \   &  $R \text{\,[mm]}  $  	&  $t \text{\,[nm]}$   	&  $W \text{\,[mm]}$  		\\ 
\hline

\textbf{Fig.\ 1b} \ \  &	Square			& 25.7			& 167			& 6.7					\\
\hline

\textbf{Fig.\ 2} \ \ 
&Ellipse			& 25.7			& 152			& 8.6						\\
&Tangential polygon	\ & 25.7			& 166			& 5.3						\\
\hline

\textbf{Fig.\ 3} \ \ &Triangle			& 34.5			& 152			& 6.5					\\
&Rectangle			& 38.6			& 162			& 9.5				\\
&Hexagon			& 25.7			& 158			& 8.4					\\
&Tangential polygon	\ & 25.7			& 166			& 5.3						\\
&Semicircle			& 34.5			& 151			& 7.7					\\
&Circle			& 34.5			& 157			& 7.7							\\
&Ellipse			& 25.7			& 152			& 8.6						\\
\hline
\hline
\end{tabular}
\vspace{1em}
\caption{
Experimental parameters for the  spherical shells ($\kappa > 0$) in the main text.  
Radius of curvature $R$, thickness $t$, and initial planar width $W$ are reported. The choice of  width depends on the shape: we use a radius, semi-major axis length, or half of a long diagonal or side as appropriate.
}
\label{tab:exp-sphere}
\end{table}

\begin{table}[H]
\centering{}\begin{tabular}{l l c c c c  }
\hline
\hline
    & Planform shape	\  &     $R \text{\,[cm]}  $  &  $t \text{\,[$\mu$m]} $  &  $W \text{\,[cm]}$  &  $K \text{\,[Pa/m]}$   		\\ 
\hline

\textbf{Fig.\ 1a} \ \   & Square ($\kappa <0$) \ & 6.4,7.8 &     1	&   		1		        & 	100			\\
\textbf{Fig.\ 1b} \ \  & Square ($\kappa >0$) \ & 20 &  5	&  			3.05	        & 	2000			\\
\hline
\textbf{Fig.\ 4a} \ \  & Rectangle ($\kappa <0$) \ & --- &  2	&  	2	        & 	100			\\
\textbf{Fig.\ 4b} \ \  & Rectangle ($\kappa >0$) \ & --- &  7	&  	3.6	        & 	2000			\\
\textbf{Fig.\ 4b} \ \  & Ellipse ($\kappa >0$) \ & --- &  7	&  			4	        & 	2000			\\
\hline
\hline
\end{tabular}
\vspace{1em}
\caption{
Parameters for the simulated shells in the main text. 
Radius of curvature $R$ (unsigned principal radii for $\kappa <0$), thickness $t$, initial width $W$, and substrate stiffness $K$ are reported.
A Young's modulus of $E=2$ MPa and Poisson's ratio of $\nu=0.495$ is used.
}
\label{tab:simulation}
\end{table}

\clearpage

\section{Supplementary Figures}\label{sec:SIfigs}


\begin{figure}[H]
\centering{}
\includegraphics[width=.99\linewidth]{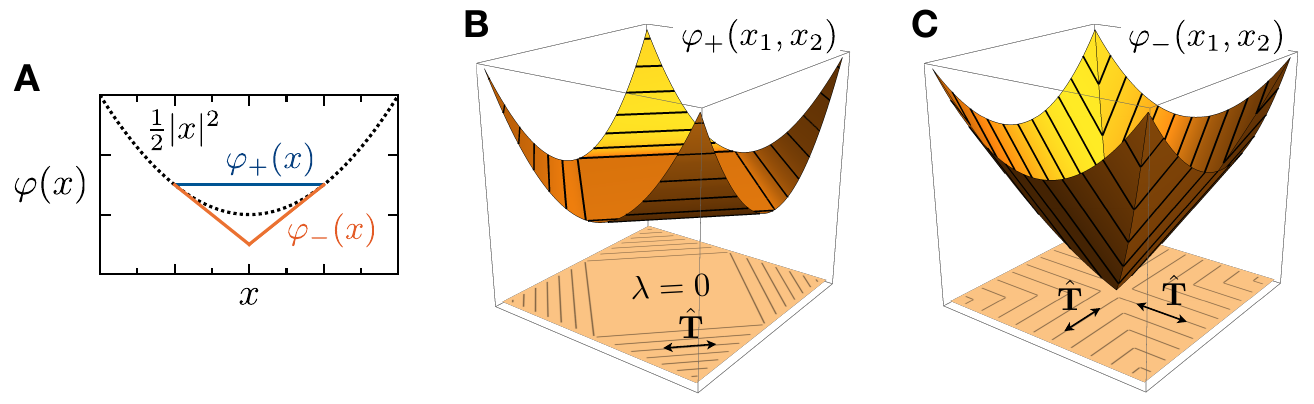}
\hypertarget{fig:AirySquare}{}
\caption{\textbf{Convex Airy potentials.} The patterns of positively and negatively curved shells are predicted using the largest and smallest convex extensions $\varphi_+$ and $\varphi_-$ of  $|\mathbf{x}|^2/2$ into the planform $\Omega$, shown here in a one-dimensional example (A), and for a positively curved square (B) and a negatively curved square (C). (A): Amongst all convex extensions,  $\varphi_+$ and $\varphi_-$ set the extremes. (B-C): Bold lines show the ruled parts of the graphs of $\varphi_+$ and $\varphi_-$. They lie above the predicted ordered domains. 
}
\label{fig:AirySquare} 
\end{figure}

\begin{figure}[H]
\centering{}
\includegraphics[width=.98\linewidth]{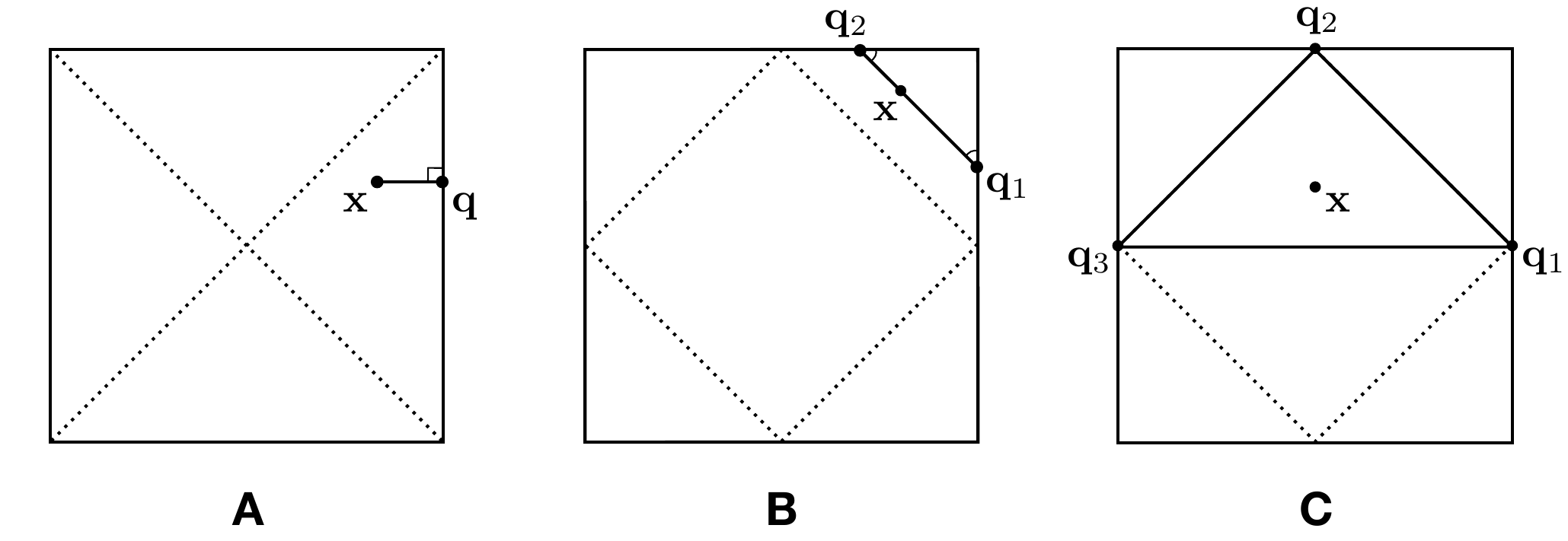}
\hypertarget{fig:bounding-schematics}{}
\caption{\textbf{Solution strategies for a square.} (A): The smallest convex extension $\varphi_-(\mathbf{x})$ is found using the closest boundary point $\mathbf{q}=\mathbf{P}_{\partial\Omega}(\mathbf{x})$ to $\mathbf{x}$. (B-C): The largest convex extension $\varphi_+(\mathbf{x})$ is found using two boundary points $\mathbf{q}_1$ and $\mathbf{q}_2$ as in (B), or three boundary points $\mathbf{q}_1$, $\mathbf{q}_2$, and $\mathbf{q}_3$ as in (C). In each, $\mathbf{x}$ belongs to the convex hull of $\{\mathbf{q}_i\}$.
}
\label{fig:bounding-schematics} 
\end{figure}

\begin{figure}[H]
\centering{}
\includegraphics[width=.4\linewidth]{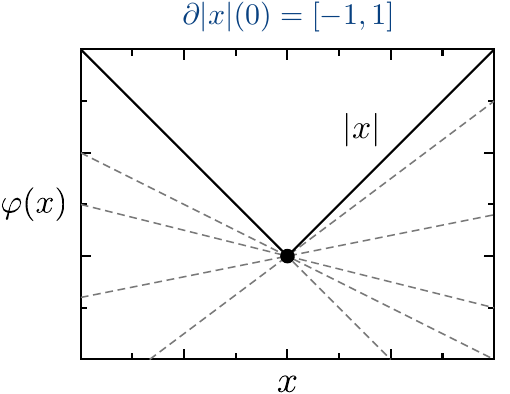}
\hypertarget{fig:subdiff-schematic}{}
\caption{\textbf{Subdifferential of a convex function.} The subdifferential $\partial|x|(0)$ of the absolute value function at $x=0$ is the interval $[-1,1]$. By definition, it contains the slopes of all tangent lines on or below its graph through $(0,0)$. We use this concept to derive our simple rules.
}
\label{fig:subdiff-schematic} 
\end{figure}

\clearpage


\section{Supplementary Text}

This Supplementary Text contains further information on the locking stress approach to confined shells.
We pick up the discussion from the Methods Theory section, where the concept of a \textit{saddle point} $(\mathbf{u}_\text{eff},\boldsymbol{\sigma}_L)$ for the Lagrangian
\begin{equation}
\mathcal{L}\{\mathbf{u}_\text{eff},\boldsymbol{\sigma}_L\}=\Delta A\{\mathbf{u}_\text{eff}\}+\int_{\Omega}\boldsymbol{\sigma}_L:\boldsymbol{\varepsilon}_\text{eff}(\mathbf{u}_\text{eff})\label{eq:Lagrangian}
\end{equation}
was defined. There, we explained why  $\mathbf{u}_\text{eff}$ solves the  Maximum Coverage Problem
\begin{equation}
\min_{\mathbf{u}_\text{eff}}\,\max_{\boldsymbol{\sigma}_L\geq\mathbf{0}}\,\mathcal{L}\left\{ \mathbf{u}_\text{eff},\boldsymbol{\sigma}_L\right\} = \min_{\boldsymbol{\varepsilon}_\text{eff}(\mathbf{u}_\text{eff})\leq\mathbf{0}}\,\Delta A\{\mathbf{u}_\text{eff}\}.
\end{equation}
We also claimed, but did not prove, that $\boldsymbol{\sigma}_L$ solves the dual problem 
\begin{equation}\label{eq:claimed-dual}
\max_{\boldsymbol{\sigma}_L\geq\mathbf{0}}\min_{\mathbf{u}_\text{eff}}\,\mathcal{L}\left\{ \mathbf{u}_\text{eff},\boldsymbol{\sigma}_L\right\} =\max_{\substack{\nabla\cdot\boldsymbol{\sigma}_L=0\text{ on }\mathbb{R}^{2}\\
\boldsymbol{\sigma}_L\geq\mathbf{0}\text{ on }\mathbb{R}^{2}\\
\boldsymbol{\sigma}_L=\mathbf{I}\text{ exterior to }\Omega
}
}\,-\frac{1}{2}\int_{\mathbb{R}^{2}}\mathbf{M}:(\boldsymbol{\sigma}_L-\mathbf{I})
\end{equation}
written using the ``misfit'' tensor $\mathbf{M}=\nabla p \otimes \nabla p$. Finally, we derived the orthogonality relation 
\begin{equation}
\boldsymbol{\sigma}_L:\boldsymbol{\varepsilon}_\text{eff}=0
\end{equation} 
for saddle points of the Lagrangian. These three facts comprise a method for predicting wrinkle domains, the details of which we go through now. 

We start in \prettyref{subsec:The-Lagrange-multipliers} by justifying the dual problem from \eqref{eq:claimed-dual}. We also show how to solve it using convex Airy potentials. In \prettyref{subsec:Worked-examples}, we apply the resulting solution formulas to find the wrinkle domains of a positively and a negatively curved square. \prettyref{subsec:Derivation-of-Rules} ends by deriving our general rule set for wrinkles from the main text. 

\medskip
\noindent\textit{Notation}. 
A symmetric matrix $\mathbf{A}$ is \textit{non-negative} (denoted $\mathbf{A}\geq \mathbf{0}$) or \textit{non-positive} (denoted $\mathbf{A}\leq \mathbf{0}$) if its eigenvalues are  non-negative  or non-positive, respectively. 
We use the inner product
$\mathbf{A}:\mathbf{B} = \sum_{i,j} A_{ij}B_{ij}$ on symmetric matrices. Note $\mathbf{A} \geq \mathbf{0}$ if and only if $\mathbf{A}:\mathbf{B} \geq 0$ for all $\mathbf{B}\geq \mathbf{0}$. 

\subsection{Solution formulas for the locking stress \label{subsec:The-Lagrange-multipliers}}

See \prettyref{subsubsec:minimax} for a proof of \eqref{eq:claimed-dual}.
See \prettyref{subsubsec:convex-Airy-potentials} for the Airy potential formulation of the dual, and \prettyref{subsubsec:abstract-solns} for a set of ``abstract solution formulas'' which we manipulate in a later section to obtain our simple rules.

\subsubsection{Derivation of the dual problem}\label{subsubsec:minimax}

Consider the max-min procedure
\begin{equation}
\max_{\boldsymbol{\sigma}_L\geq\mathbf{0}}\min_{\mathbf{u}_\text{eff}}\,\mathcal{L}\left\{ \mathbf{u}_\text{eff},\boldsymbol{\sigma}_L\right\} \label{eq:max-min}
\end{equation}
which characterizes the $\boldsymbol{\sigma}_L$ part of a saddle point for $\mathcal{L}$. 
As the inner minimization over $\mathbf{u}_\text{eff}$ is unconstrained, we can evaluate it by setting $0=\frac{\delta\mathcal{L}}{\delta \mathbf{u}_\text{eff}}$. This gives
\begin{equation}
0=-\int_{\partial\Omega}\delta \mathbf{u}_\text{eff}\cdot\hat{\mathbf{n}}\,ds+\int_{\Omega}\boldsymbol{\sigma}_L:\nabla\delta \mathbf{u}_\text{eff}\label{eq:weakly-div-free}
\end{equation}
for all $\delta \mathbf{u}_\text{eff}$, a condition we shall simplify using the divergence theorem. To do so, we first extend $\boldsymbol{\sigma}_L$ to the exterior of $\Omega$ by taking
\begin{equation}\label{eq:extended-bcs}
\boldsymbol{\sigma}_L(\mathbf{x})=\mathbf{I} \quad\text{for }\mathbf{x}\notin \overline{\Omega}
\end{equation}
where $\overline{\Omega}$ stands for the closure of $\Omega$. Then, \eqref{eq:weakly-div-free} rearranges to say that \begin{equation}
0=\int_{\mathbb{R}^{2}}\boldsymbol{\sigma}_L:\nabla\delta \mathbf{u}_\text{eff} 
\end{equation}
for all $\delta \mathbf{u}_\text{eff}$, i.e.,  $\boldsymbol{\sigma}_L$ is \textit{weakly divergence-free} as a tensor field on $\mathbb{R}^2$. Note this includes a matching condition  at the boundary of $\Omega$, which allows for the possibility that $\boldsymbol{\sigma}_L\hat{\mathbf{n}}\neq\hat{\mathbf{n}}$ at $\partial\Omega$  when viewed from its interior. The condition \eqref{eq:extended-bcs} is simply a way of enforcing the usual boundary conditions $\boldsymbol{\sigma}_L\hat{\mathbf{n}}=\hat{\mathbf{n}}$ at $\partial\Omega$ from the exterior. This ``relaxation procedure'' turns out to be necessary to ensure the existence of a maximizing $\boldsymbol{\sigma}_L$, especially in the positively curved case \cite{tobasco2021curvature}. 

Now if $\boldsymbol{\sigma}_L$ is not weakly divergence-free, it must be that $\min_{\mathbf{u}_\text{eff}}\,\mathcal{L}=-\infty$. Otherwise, the minimum is finite and evaluating it produces the desired dual. 
To see this, apply 
\prettyref{eq:weakly-div-free} with  $\delta \mathbf{u}_\text{eff}=\mathbf{u}$ in the case that $\nabla\cdot\boldsymbol{\sigma}_L=0$. The result is that 
\begin{align*}
\mathcal{L}\{\mathbf{u}_\text{eff},\boldsymbol{\sigma}_L\} & =\int_{\Omega}\frac{1}{2}\text{tr}\,\mathbf{M}\,d\mathbf{x}-\int_{\partial\Omega}\mathbf{u}_\text{eff}\cdot\hat{\mathbf{n}}\,ds+\int_{\Omega}\boldsymbol{\sigma}_L:\left(\nabla \mathbf{u}_\text{eff}-\frac{1}{2}\mathbf{M}\right)\\
 & =\int_{\mathbb{R}^2}-\frac{1}{2}\mathbf{M}:\left(\boldsymbol{\sigma}_L-\mathbf{I}\right)
\end{align*}
which is clearly independent of $\mathbf{u}_\text{eff}$. We have arrived at the dual problem for the locking stress:
\begin{equation}\label{eq:dual-problem-derived}
\max_{\substack{\nabla\cdot\boldsymbol{\sigma}_L=0\text{ on }\mathbb{R}^{2}\\
\boldsymbol{\sigma}_L\geq\mathbf{0}\text{ on }\mathbb{R}^{2}\\
\boldsymbol{\sigma}_L=\mathbf{I}\text{ exterior to }\Omega
}
}\,\int_{\mathbb{R}^2}-\frac{1}{2}\mathbf{M}:(\boldsymbol{\sigma}_L-\mathbf{I}).
\end{equation}


\subsubsection{Convex Airy potentials}\label{subsubsec:convex-Airy-potentials}

Having recovered the dual problem for $\boldsymbol{\sigma}_L$, we proceed to rewrite it using an Airy-like potential we call $\varphi$. Introduce the rotated gradient $\nabla^\perp = (-\partial_2 , \partial _1)$ and make the substitution 
\begin{equation}
\boldsymbol{\sigma}_L=\nabla^{\perp}\nabla^{\perp}\varphi\quad\text{where}\quad\nabla^{\perp}\nabla^{\perp}=\left(\begin{array}{cc}
\partial_{22} & -\partial_{12}\\
-\partial_{21} & \partial_{11}
\end{array}\right)\label{eq:grad-grad-notation}
\end{equation}
for some to be determined function $\varphi(\mathbf{x})$. While the ``rotated Hessian'' notation $\nabla^\perp\nabla^\perp$ is not 
quite standard, we find it speeds up the required manipulations.
For instance, since 
$\nabla \cdot \nabla^{\perp}=0$,
any $\boldsymbol{\sigma}_L$ obtained via \prettyref{eq:grad-grad-notation}
is automatically (weakly) divergence-free. 
The change of variables from $\boldsymbol{\sigma}_L$ to $\varphi$  is often referred to as introducing an Airy potential. Our potential functions depart from the usual ones due to their constraints (and, importantly, they can fail to be smooth). 

Looking back at \eqref{eq:dual-problem-derived}, we see that in order for $\boldsymbol{\sigma}_L$ to be admissible, $\varphi$ must obey 
\[
\nabla\nabla\varphi\geq\mathbf{0}\quad\text{in }\mathbb{R}^2 \quad\text{and}\quad \nabla\nabla\varphi=\mathbf{I} \quad\text{in } \mathbb{R}^2\backslash\overline{\Omega}.
\]
The first condition holds as $\nabla^\perp\nabla^\perp \varphi$ and $\nabla\nabla \varphi$ share the same eigenvalues. Thus,  $\varphi$ is a convex function with
\[
\varphi(\mathbf{x})=\frac{1}{2}|\mathbf{x}|^2 + a(\mathbf{x})\quad \text{where} \quad \nabla \nabla a(\mathbf{x}) = 0 \quad\text{for}\quad \mathbf{x}\notin\Omega.
\]
If $\Omega$ is \textit{simply connected} so that it has no holes, we may take $a = 0$ without loss of generality. 
Otherwise, there is the possibility that  $a$ is a different affine function within each hole (see \cite{tobasco2021curvature} for more on this case). 
Proceeding with simply connected shells, we have shown that the set of admissible $\boldsymbol{\sigma}_L$ for the dual corresponds to the set of convex Airy potentials $\varphi$ defined on $\mathbb{R}^2$ and that equal $|\mathbf{x}|^2/2$ for $\mathbf{x}\notin \Omega$. We  call such  functions \emph{convex extensions} of $|\mathbf{x}|^2/2$ into $\Omega$.

The next step is to rewrite the objective functional in \eqref{eq:dual-problem-derived} in terms of $\varphi$. A straightforward calculation using the shallow shell approximation
\[
\kappa=\det \nabla \nabla p
\]
for the initial Gaussian curvature along with our choice to call $\mathbf{M}=\nabla p\otimes \nabla p$ shows that
\[
\kappa =-\frac{1}{2}\nabla \times \nabla \times\,\mathbf{M}\quad\text{where}\quad\nabla \times \nabla \times\,\mathbf{M}=\partial_{11}M_{22}+\partial_{22}M_{11}-2\partial_{12}M_{12}.
\]
Integrating by parts twice
, we find that
\begin{align*}
\int_{\mathbb{R}^2}-\frac{1}{2}\mathbf{M}:(\boldsymbol{\sigma}_L-\mathbf{I}) & =\int_{\mathbb{R}^2}-\frac{1}{2}\mathbf{M}:\nabla^{\perp}\nabla^{\perp}\left(\varphi-\frac{|\mathbf{x}|^2}{2}\right)\\
 & =\int_{\mathbb{R}^2}-\frac{1}{2}\nabla \times \nabla \times\,\mathbf{M}\left(\varphi-\frac{|\mathbf{x}|^2}{2}\right)\\
 &=\int_{\Omega}\left(\varphi-\frac{|\mathbf{x}|^2}{2}\right)\kappa
\end{align*}
where in the last step we used that $\varphi = |\mathbf{x}|^2/2$ outside $\Omega$. Maximizing this last integral amongst all admissible convex extensions produces the sought after locking stress $\boldsymbol{\sigma}_L$ in terms of the initial planform $\Omega$ and the initial Gaussian curvature $\kappa$. 


\subsubsection{Abstract solution formulas\label{subsubsec:abstract-solns}}

The prior sections obtained the dual problem for $\boldsymbol{\sigma}_L$ and found its Airy potential form
\begin{equation}
\max_{\substack{\varphi\text{ is convex}\\
\varphi = \frac{1}{2}|\mathbf{x}|^2\text{ outside } \Omega
}
}\,\int_{\Omega}\left(\varphi-\frac{|\mathbf{x}|^2}{2} \right)\kappa. \label{eq:dual-pblm-Airyform}
\end{equation}
We now show how to solve these problems if the initial Gaussian curvature $\kappa(\mathbf{x})$ is of one sign.  
We claim that  
\prettyref{eq:dual-pblm-Airyform} is solved by the functions
\begin{equation}
\varphi_{+}(\mathbf{x}):=\max_{\varphi}\,\varphi(\mathbf{x})\quad\text{and}\quad\varphi_{-}(\mathbf{x}):=\min_{\varphi}\,\varphi(\mathbf{x}).\label{eq:abstract-optimization}
\end{equation}
The $+$ one applies in the positively curved case $\kappa>0$, and the $-$ one applies in the negatively curved case $\kappa<0$. The proof is actually quite simple: if $\kappa> 0$ then the integral in \eqref{eq:dual-pblm-Airyform} is made largest by driving $\varphi(\mathbf{x})$ up to its maximum value; if $\kappa< 0$, one must instead make  $\varphi(\mathbf{x})$ as small as possible. What is being optimized is the choice of convex extension $\varphi$ of $|\mathbf{x}|^2/2$ into $\Omega$. Undoing the change of variables  yields 
\begin{equation}\label{eq:locking-stress-abstract-formulas}
\boldsymbol{\sigma}_L(\mathbf{x})=\begin{cases}
\nabla^{\perp}\nabla^{\perp}\varphi_{-}(\mathbf{x}) & \text{if }\kappa< 0\\
\nabla^{\perp}\nabla^{\perp}\varphi_{+}(\mathbf{x}) & \text{if }\kappa> 0
\end{cases}
\end{equation}
for the locking stress. 

At this stage, the above solution formulas may seem rather abstract. Fortunately, they can be simplified to reveal our simple rules. To prepare, we now record some partial simplifications of these formulas obtained in \cite{tobasco2021curvature}. The reader who would prefer to first go through some concrete examples may safely skip forward to \prettyref{subsec:Worked-examples} and come back as needed later on.





The first simplification concerns the smallest convex extension $\varphi_-$: given $\mathbf{x}\in \Omega$,
\begin{equation}
\varphi_{-}(\mathbf{x})=\frac{1}{2}|\mathbf{x}|^{2}-\frac{1}{2}d_{\partial\Omega}^{2}(\mathbf{x})\quad\text{where}\quad d_{\partial\Omega}(\mathbf{x})=\min_{\mathbf{q}\in\partial\Omega}\,|\mathbf{x}-\mathbf{q}|.\label{eq:negative-curvature-solution}
\end{equation}
Behind this is the elementary fact that every convex function is bounded below by its tangent planes.
We give a detailed explanation in \prettyref{subsubsec:Negatively-curved-squares} for  a square. The same reasoning leads to the general result. 

Regarding the largest convex extension $\varphi_+$: given $\mathbf{x}\in \Omega$,
\begin{equation}
\varphi_{+}(\mathbf{x})=\min_{\substack{\{\mathbf{q}_{i}\}\subset\partial\Omega\\
\{\theta_{i}\}\subset[0,1]
}
}\,\sum_i\theta_{i}\frac{1}{2}|\mathbf{q}_{i}|^{2}\quad\text{where}\quad\sum_{i}\theta_{i}\mathbf{q}_{i}=\mathbf{x}.\label{eq:positive-curvature-solution}
\end{equation}
The minimization is over all ways of writing $\mathbf{x}$ as a convex
combination of boundary points ${\mathbf{q}_i}$. At most three boundary points are required, the choice of which depends on $\mathbf{x}$. Behind this is the fact that $\varphi_{+}$ is the \textit{convex
hull} of 
\[
h(\mathbf{x})=\begin{cases}
\infty & \mathbf{x}\in\Omega\\
\frac{1}{2}|\mathbf{x}|^2 & \mathbf{x}\notin\Omega
\end{cases},
\]
i.e., the largest convex function whose graph lies on or below that of $h$. See \prettyref{subsec:Derivation-of-Rules} for a brief introduction to convex functions.

\subsection{Two worked examples\label{subsec:Worked-examples}}

Here, we demonstrate the theory on positively and negatively curved squares. To be sure, one could appeal to our general rules, which govern the patterns of curved squares in addition to many other shells (see \prettyref{subsec:Derivation-of-Rules}). Instead, our present goal is to show how it is possible to directly evaluate the solution formulas from \prettyref{subsubsec:abstract-solns}, in the case of a square reference domain:
\begin{equation}\label{eq:square-ref-domain}
\Omega=\left\{ (x_{1},x_{2}):|x_{1}|\leq1,|x_{2}|\leq1\right\}.
\end{equation}


\subsubsection{Negatively curved squares\label{subsubsec:Negatively-curved-squares}}

The patterns of a negatively curved square are given by the smallest convex extension $\varphi_{-}$ of $|\mathbf{x}|^2/2$ into the square. The question is: how can we evaluate $\varphi_-$? First, we show how to bound an arbitrary convex extension $\varphi$ from below, obtaining the inequality
\begin{equation}
\varphi(\mathbf{x})\geq\mathfrak{L}(\mathbf{x})\label{eq:desired-bound-L}
\end{equation}
for some ``bounding function'' $\mathfrak{L}(\mathbf{x})$. Different
procedures result in different bounds $\mathfrak{L}$. 
The optimal procedure is characterized by the fact that its $\mathfrak{L}$
 is itself a convex extension, giving the identity
\begin{equation}
\varphi_{-}(\mathbf{x})=\mathfrak{L}(\mathbf{x}).\label{eq:desired-equality-L}
\end{equation}
Panel (a) of \prettyref{fig:AirySquare}a illustrates our general strategy, and panel (c) shows the result for a square.
We pursue a similar strategy to recover $\varphi_{+}$ in \prettyref{subsubsec:Negatively-curved-squares}.

Begin with an arbitrary convex extension  $\varphi$ of $|\mathbf{x}|^2/2$
into $\Omega$. Let $\mathbf{x}\in\Omega$ and let $\mathbf{q}\in\partial\Omega$ be
a point that is closest to $\mathbf{x}$ amongst all boundary points, as in \prettyref{fig:bounding-schematics}a. 
Parameterize the segment $\mathbf{x}\mathbf{q}$ as $(1-t)\mathbf{x}+t\mathbf{q}$ for $t\in[0,1]$, and let 
\[
f(t)=\varphi\left((1-t)\mathbf{x}+t\mathbf{q}\right).
\]
Expanding in a Taylor series around $t=1$, we get that
\[
f(t)=f(1)-f'(1)(1-t)-\int_{s=t}^{s=1}\frac{f''(s)}{2}(t-s)\,ds
\]
for $t\in[0,1]$. By the boundary conditions for $\varphi$, 
\[
f(1)=\varphi(\mathbf{q})=\frac{1}{2}|\mathbf{q}|^{2}\quad\text{and}\quad f'(1)=(\mathbf{q}-\mathbf{x})\cdot \mathbf{q}.
\]
Since $\varphi$ is convex, so is $f$ and therefore 
\[
f''(t)\geq0.
\]
From these observations follows the inequality
\[
f(t)\geq\frac{1}{2}|\mathbf{q}|^{2}-(\mathbf{q}-\mathbf{x})\cdot \mathbf{q}(1-t).
\]
Taking $t=0$ there results
\[
\varphi(\mathbf{x})\geq\frac{1}{2}|\mathbf{q}|^{2}-(\mathbf{q}-\mathbf{x})\cdot \mathbf{q}=\mathbf{x}\cdot \mathbf{q}-\frac{1}{2}|\mathbf{q}|^{2}.
\]
Note $\mathbf{q} = \mathbf{P}_{\partial\Omega}(\mathbf{x})$, where $\mathbf{P}_{\partial\Omega}(\mathbf{x})$ is the closest boundary point to $\mathbf{x}$, as in the main text. 

So far, we have shown that an arbitrary convex extension $\varphi$ satisfies
\[
\varphi(\mathbf{x})\geq \mathfrak{L}(\mathbf{x}):= \mathbf{x}\cdot \mathbf{P}_{\partial\Omega}(\mathbf{x})-\frac{1}{2}|\mathbf{P}_{\partial\Omega}(\mathbf{x})|^{2}
\]
for any $\mathbf{x}\in \Omega$. 
In fact, the same lower bound holds in general (we have yet to use that the reference domain is a square). For the square $\Omega$ in  \prettyref{eq:square-ref-domain},
\begin{equation}
\mathbf{P}_{\partial\Omega}(\mathbf{x})=\begin{cases}
(1,x_{2}) & x_{1}>0,-x_{1}<x_{2}<x_{1}\\
(x_{1},1) & x_{2}>0,-x_{2}<x_{1}<x_{2}\\
(-1,x_{2}) & x_{1}<0,x_{1}<x_{2}<-x_{1}\\
(x_{1},-1) & x_{2}<0,x_{2}<x_{1}<-x_{2}
\end{cases}.\label{eq:defn-of-L2}
\end{equation}
This proves the
anticipated bound \prettyref{eq:desired-bound-L}, where by default  $\mathfrak{L}(\mathbf{x})=|\mathbf{x}|^{2}/2$ for $\mathbf{x}\notin\Omega$. 

Next, we show that $\varphi_- = \mathfrak{L}$. To do this, we check that the bounding function
\[
\mathfrak{L}(\mathbf{x})=\begin{cases}
\mathbf{x}\cdot \mathbf{P}_{\partial\Omega}(\mathbf{x})-\frac{1}{2}|\mathbf{P}_{\partial\Omega}(\mathbf{x})|^{2} & \mathbf{x}\in\Omega\\
\frac{1}{2}|\mathbf{x}|^{2} & \mathbf{x}\notin\Omega
\end{cases}
\]
from above is itself a convex extension of $|\mathbf{x}|^2/2$ into $\Omega$. Evidently $\mathfrak{L}(\mathbf{x}) = |\mathbf{x}|^2/2$ outside $\Omega$, so we  only need to show it is convex. 
Given $\mathbf{x}\in\Omega$, note that
\[
\left(\mathbf{P}_{\partial\Omega}(\mathbf{x})-\mathbf{x}\right)\cdot\nabla \mathbf{P}_{\partial\Omega}(\mathbf{x})=0
\]
as the closest boundary point to $\mathbf{x}$ is constant along the 
segment between $\mathbf{x}$ and $\mathbf{P}_{\partial\Omega}(\mathbf{x})$. So, 
\[
\nabla\mathfrak{L}(\mathbf{x})=\mathbf{P}_{\partial\Omega}(\mathbf{x}).
\]
Differentiating once more and using \prettyref{eq:defn-of-L2},
we get for $\mathbf{x}\in\Omega$ that
\[
\nabla\nabla\mathfrak{L}(\mathbf{x})=\begin{cases}
(0,1)\otimes(0,1) & x_{1}>0,-x_{1}<x_{2}<x_{1}\\
(-1,0)\otimes(-1,0) & x_{2}>0,-x_{2}<x_{1}<x_{2}\\
(0,-1)\otimes(0,-1) & x_{1}<0,x_{1}<x_{2}<-x_{1}\\
(1,0)\otimes(1,0) & x_{2}<0,x_{2}<x_{1}<-x_{2}
\end{cases}.
\]
Since $\mathbf{a}\otimes \mathbf{a} =(-\mathbf{a})\otimes (-\mathbf{a})$, the signs of the vectors in the outer products above are immaterial. What matters is that the eigenvalues of $\nabla\nabla \mathfrak{L}$ are non-negative. They are zero and one, so $\mathfrak{L}$ is indeed convex in the four regions identified above. One can also check that $\mathfrak{L}$ is convex at their boundaries, as well as at  $\partial\Omega$. See \prettyref{fig:AirySquare}c.

Combining the lower bound $\varphi\geq\mathfrak{L}$ and the admissibility 
of $\mathfrak{L}$, we conclude that 
\[
\varphi_{-}(\mathbf{x})=\mathfrak{L}(\mathbf{x})
\]
for all $\mathbf{x}$. 
Applying the solution formula \eqref{eq:locking-stress-abstract-formulas}, we get that
\[
\boldsymbol{\sigma}_L(\mathbf{x})=\begin{cases}
(-1,0)\otimes(-1,0) & x_{1}>0,-x_{1}<x_{2}<x_{1}\\
(0,-1)\otimes(0,-1) & x_{2}>0,-x_{2}<x_{1}<x_{2}\\
(1,0)\otimes(1,0) & x_{1}<0,x_{1}<x_{2}<-x_{1}\\
(0,1)\otimes(0,1) & x_{2}<0,x_{2}<x_{1}<-x_{2}
\end{cases}.
\]
By the argument in the main text, the ordered wrinkles are therefore shown to have their peaks and troughs along the unit vector field
\[
\hat{\mathbf{T}}(\mathbf{x})=\begin{cases}
(-1,0) & x_{1}>0,-x_{1}<x_{2}<x_{1}\\
(0,-1) & x_{2}>0,-x_{2}<x_{1}<x_{2}\\
(1,0) & x_{1}<0,x_{1}<x_{2}<-x_{1}\\
(0,1) & x_{2}<0,x_{2}<x_{1}<-x_{2}
\end{cases}.
\]
The reader should compare this result against the patterns in Fig.~1a of the main text. 


\subsubsection{Positively curved squares\label{subsubsec:Positively-curved-squares}}

Next, we solve for the wrinkles of a positively curved square. This time, we must find the largest convex extension $\varphi_{+}$ of $|\mathbf{x}|^2/2$ into $\Omega$. We proceed as above: first, we show a general
upper bound
\begin{equation}
\varphi(\mathbf{x})\leq\mathfrak{U}(\mathbf{x})\label{eq:desired-bound-U}
\end{equation}
holding for all convex extensions $\varphi$. Then, we check that $\mathfrak{U}(\mathbf{x})$ is itself a convex extension, so that
\begin{equation}
\varphi_{+}(\mathbf{x})=\mathfrak{U}(\mathbf{x}).\label{eq:desired-equality-U}
\end{equation}
The reader may wish to refer to \prettyref{fig:AirySquare}a,b in what follows. 

Let $\varphi$ be a convex extension of $|\mathbf{x}|^2/2$
into $\Omega$. Decompose $\Omega$ into five parts: the
isosceles triangles
\begin{align*}
T_{\text{ne}} & =\left\{ (x_{1},x_{2}):x_{1}+x_{2}>1,x_{1}<1,x_{2}<1\right\} \\
T_{\text{nw}} & =\left\{ (x_{1},x_{2}):-x_{1}+x_{2}>1,x_{1}>-1,x_{2}<1\right\} \\
T_{\text{sw}} & =\left\{ (x_{1},x_{2}):x_{1}+x_{2}<-1,x_{1}>-1,x_{2}>-1\right\} \\
T_{\text{se}} & =\left\{ (x_{1},x_{2}):-x_{1}+x_{2}<-1,x_{1}<1,x_{2}>-1\right\} 
\end{align*}
and the leftover diamond
\[
D=\left\{ (x_{1},x_{2}):|x_{1}|+|x_{2}|\leq1\right\} .
\]
We obtain an upper bound in each part. First, let $\mathbf{x}\in T_{\text{ne}}$. Abusing notation slightly, we note it belongs to a unique line segment $\mathbf{q}_{1}\mathbf{q}_{2}$ perpendicular to $(1,1)$ with $\mathbf{q}_{1},\mathbf{q}_{2}\in\partial\Omega$. 
That is,
\[
\mathbf{x}=\theta \mathbf{q}_{1}+(1-\theta)\mathbf{q}_{2}\quad\text{where}\quad \theta \in [0,1].
\]
In particular, we can take 
\[
\mathbf{q}_{1}=(1,x_{1}+x_{2}-1)\quad\text{and}\quad \mathbf{q}_{2}=(x_{1}+x_{2}-1,1)
\]
as in \prettyref{fig:bounding-schematics}b. Define the function
\[
f(\theta)=\varphi\left(\theta \mathbf{q}_{1}+(1-\theta) \mathbf{q}_{2}\right).
\]
The boundary conditions for $\varphi$ require that
\[
f(0)=\varphi(\mathbf{q}_{1})=\frac{1}{2}|\mathbf{q}_{1}|^{2}\quad\text{and}\quad f(1)=\varphi(\mathbf{q}_{2})=\frac{1}{2}|\mathbf{q}_{2}|^{2}.
\]
Since $\varphi$ is convex, so is $f$. Hence,
\[
f(\theta)\leq\theta f(0)+(1-\theta)f(1)
\]
for all $\theta\in[0,1]$. Unpacking the definitions, we conclude
that
\[
\varphi(\mathbf{x})\leq\theta\frac{|\mathbf{q}_{1}|^{2}}{2}+(1-\theta)\frac{|\mathbf{q}_{2}|^{2}}{2}=\frac{1}{2}\left(1+(x_{1}+x_{2}-1)^{2}\right)
\]
for all $\mathbf{x}\in T_{\text{ne}}$. This is the desired upper bound \prettyref{eq:desired-bound-U} in the triangle
$T_{\text{ne}}$. The remaining triangles $T_{\text{nw}}$, $T_{\text{sw}}$, and $T_{\text{se}}$ can be
dealt with similarly. 

If instead $\mathbf{x}\in D$, it is a barycenter of a triangle $\mathbf{q}_{1}\mathbf{q}_{2}\mathbf{q}_{3}$ with $\mathbf{q}_{1},\mathbf{q}_{2},\mathbf{q}_{3}\in\partial\Omega\cap \partial D$. That is,
\[
\mathbf{x}=\sum_{i=1}^{3}\theta_{i}\mathbf{q}_{i}\quad\text{where}\quad\sum_{i=1}^{3}\theta_{i}=1\quad\text{and}\quad \theta_{i}\in[0,1]
\]
and where each $\mathbf{q}_i$ is a contact point of the inscribed circle to the
square. The setup is as in \prettyref{fig:bounding-schematics}c. By its definition,
\[
\varphi(\mathbf{q}_{i})=\frac{1}{2}|\mathbf{q}_{i}|^{2}=\frac{1}{2}
\]
for each $i$. It follows that
\[
\varphi(\mathbf{x})\leq\sum_{i=1}^{3}\theta_{i}\frac{1}{2}|\mathbf{q}_{i}|^{2}=\frac{1}{2}
\]
by convexity. We have proved the upper bound
\[
\varphi(\mathbf{x})\leq \mathfrak{U}(\mathbf{x}):=\begin{cases}
\frac{1}{2}|\mathbf{x}|^2 & \mathbf{x}\notin \Omega\\
\frac{1}{2}\left(1+(x_{1}+x_{2}-1)^{2}\right) & \mathbf{x}\in T_{\text{ne}}\\
\frac{1}{2}\left(1+(-x_{1}+x_{2}-1)^{2}\right) & \mathbf{x}\in T_{\text{nw}}\\
\frac{1}{2}\left(1+(-x_{1}-x_{2}-1)^{2}\right) & \mathbf{x}\in T_{\text{sw}}\\
\frac{1}{2}\left(1+(x_{1}-x_{2}-1)^{2}\right) & \mathbf{x}\in T_{\text{se}}\\
\frac{1}{2} & \mathbf{x}\in D
\end{cases}
\]
anticipated in \prettyref{eq:desired-bound-U}. Note 
$\mathfrak{U}(\mathbf{x})=|\mathbf{x}|^{2}/2$ by default for $\mathbf{x}\notin\Omega$. 

The next step is to check that $\mathfrak{U}$ is a convex extension of $|\mathbf{x}|^2/2$
into $\Omega$. It is obviously equal to $|\mathbf{x}|^2/2$ outside $\Omega$, so we only need to verify it is convex. 
For $\mathbf{x}\in\Omega$, we see that
\[
\nabla\nabla\mathfrak{U}(\mathbf{x})=\begin{cases}
(1,1)\otimes(1,1) & \mathbf{x}\in T_{\text{ne}}\\
(-1,1)\otimes(-1,1) & \mathbf{x}\in T_{\text{nw}}\\
(-1,-1)\otimes(-1,-1) & \mathbf{x}\in T_{\text{sw}}\\
(1,-1)\otimes(1,-1) & \mathbf{x}\in T_{\text{se}}\\
0 & \mathbf{x}\in D
\end{cases}.
\]
Note $\nabla\nabla\mathfrak{U}(\mathbf{x})$ either has one non-negative eigenvalue, or is
equal to zero. In any case, $\nabla\nabla\mathfrak{U}\geq0$ so that $\mathfrak{U}$
is convex. Again, one must be a bit more careful to check its convexity at
$\partial\Omega$, but this is not so difficult to do, especially with \prettyref{fig:AirySquare}b in mind.

Having shown the upper bound $\varphi\leq\mathfrak{U}$ and concluded
that $\mathfrak{U}$ is convex, we get that
\[
\varphi_{+}(\mathbf{x})=\mathfrak{U}(\mathbf{x})
\]
for all $\mathbf{x}$. By \eqref{eq:locking-stress-abstract-formulas},
\[
\boldsymbol{\sigma}_L(\mathbf{x})=\begin{cases}
(-1,1)\otimes(-1,1) & \mathbf{x}\in T_{\text{ne}}\\
(-1,-1)\otimes(-1,-1) & \mathbf{x}\in T_{\text{nw}}\\
(1,-1)\otimes(1,-1) & \mathbf{x}\in T_{\text{sw}}\\
(1,1)\otimes(1,1) & \mathbf{x}\in T_{\text{se}}\\
0 & \mathbf{x}\in D
\end{cases}.
\]
The ordered wrinkles of a positively
curved square are therefore shown to have their peaks and troughs along the unit vector field
\[
\hat{\mathbf{T}}(\mathbf{x})=\begin{cases}
(-\frac{1}{\sqrt{2}},\frac{1}{\sqrt{2}}) & \mathbf{x}\in T_{\text{ne}}\\
(-\frac{1}{\sqrt{2}},-\frac{1}{\sqrt{2}}) & \mathbf{x}\in T_{\text{nw}}\\
(\frac{1}{\sqrt{2}},-\frac{1}{\sqrt{2}}) & \mathbf{x}\in T_{\text{sw}}\\
(\frac{1}{\sqrt{2}},\frac{1}{\sqrt{2}}) & \mathbf{x}\in T_{\text{se}}
\end{cases}.
\]
As $\boldsymbol{\sigma}_L=\mathbf{0}$ in the diamond $D$, it does not dictate a preferred wrinkle direction there. The reader should compare this result against the patterns in Fig.~1b of the main text. 

\subsection{Derivation of the simple rules\label{subsec:Derivation-of-Rules}}

The previous two sections explained how to solve for the locking stress of a given shell, first in the abstract in \prettyref{subsec:The-Lagrange-multipliers} and then in two concrete examples (curved squares) in \prettyref{subsec:Worked-examples}. 
In this final section, we apply our framework to deduce the following set of general rules:
\begin{enumerate}
\item[(i)] If $\kappa<0$, the shell's ordered wrinkles fall along segments of quickest exit from its planform $\Omega$. Such paths are line segments that meet the boundary $\partial\Omega$ perpendicularly, and meet each other at
the medial axis $\mathfrak{M}$ of $\Omega$. 
\item[(ii)] If $\kappa>0$, the shell's ordered wrinkles fall along a family of line segments connecting pairs of boundary points $\{\mathbf{q},\mathbf{r}\}\subset\partial\Omega$. Each such pair arises as the unique closest boundary points to some $\mathbf{p}\in \mathfrak{M}$. 
\item[(iii)] If $\kappa>0$, any disorder that occurs resides within a family of convex regions, constructed using the points on the medial axis leftover from the previous rule. Each such region is the convex hull of three or more closest boundary points to some $\mathbf{p}\in\mathfrak{M}$. \end{enumerate}
Rules (i) and (ii) give one of the two reciprocal pairings from the main text: the ordered wrinkles of negatively and positively curved shells pair via isoceles triangles. Rule (iii) is the other one: it pairs the disordered regions of positively curved shells with the ordered wrinkles of their negatively curved counterparts. Recall the \textit{convex hull} of a subset $Q\subset \mathbb{R}^2$ (denoted $\text{co }Q$) is the smallest subset of $\mathbb{R}^2$ that both contains $Q$, and has the property that if it contains the points $\mathbf{q}$ and $\mathbf{r}$ then it contains the line segment $\mathbf{q}\mathbf{r}$.  

The main point of our proof is to establish the following formulas for the locking stress, which apply for any simply connected shell whose initial Gaussian curvature $\kappa(\mathbf{x})$ is of one sign:
\begin{equation}
\boldsymbol{\sigma}_L(\mathbf{x})=\begin{cases}
\mathbf{R}^{T}\nabla \mathbf{P}_{\partial\Omega}(\mathbf{x})\mathbf{R} & \text{if }\kappa< 0\\
\mathbf{R}^{T}\nabla \mathbf{P}_{\mathfrak{M}}(\mathbf{x})\mathbf{R} & \text{if }\kappa> 0
\end{cases}. \label{eq:formulas-for-sigmaL}
\end{equation}
Here, $\mathbf{P}_{\partial\Omega}$ and $\mathbf{P}_{\mathfrak{M}}$ are as in Fig.~5 of the main text (see \prettyref{subsubsec:Identifying-the-gradients} for their precise definitions). The matrix $\mathbf{R}$ is a ninety-degree counterclockwise rotation. 
We break the argument into three steps across Sections \ref{subsubsec:Legendre-transforms}-\ref{subsubsec:Deriving-the-multipliers}. Rules (i)-(iii) follow at the very end. Before starting the proof, the reader may wish to review the abstract solution formulas from  \prettyref{subsubsec:abstract-solns}, and especially their partial simplifications   \prettyref{eq:negative-curvature-solution} and \prettyref{eq:positive-curvature-solution}. \prettyref{subsubsec:primer-convexity} is a primer on convex functions. 

\subsubsection{A primer on convex functions\label{subsubsec:primer-convexity}}

The following facts about convex functions will be used (see, e.g., \cite{rockafellar1970convex,dacorogna2008direct} for their proofs). A function $\varphi$ defined on $\mathbb{R}^n$ is said to be  \emph{convex} if
\[
\varphi(\theta \mathbf{x}+(1-\theta)\mathbf{y})\leq\theta\varphi(\mathbf{x})+(1-\theta)\varphi(\mathbf{y})
\]
for all $\mathbf{x},\mathbf{y}\in\mathbb{R}^n$ and $\theta\in[0,1]$. Geometrically, this says that the graph of $\varphi$ lies on or below any line segment
connecting two of its points. For smooth $\varphi$, convexity is equivalent to the condition that $\nabla\nabla\varphi\geq0$, meaning its eigenvalues are non-negative. In fact, this equivalence holds more generally, provided the second derivatives of $\varphi$ are interpreted in a suitable (distributional) sense. 

While a general convex function need not be smooth, it will be differentiable almost everywhere, i.e., the set of points where $\nabla\varphi$ fails to exist is of measure zero when $\varphi$ is convex. Even if a convex function fails to be differentiable, one can still define the notion of a ``tangent plane'' to its graph. There are simply multiple choices of tangent planes at points of non-differentiability. Following the usual convention, we refer to those tangent planes that lie on or below the given graph. 
Correspondingly, the \emph{subdifferential} of $\varphi$ at $\mathbf{x}$ is defined as the set
\[
\partial\varphi(\mathbf{x})=\left\{ \mathbf{p}\in\mathbb{R}^2:\varphi(\mathbf{y})\geq \varphi(\mathbf{x}) + \mathbf{p}\cdot(\mathbf{y}-\mathbf{x})\quad\text{for all }\mathbf{y}\in\mathbb{R}^2\right\}. 
\]
Points $\mathbf{p}\in \partial\varphi(\mathbf{x})$ correspond to tangent planes. 
Conveniently, $\varphi$ is differentiable at $\mathbf{x}$ in the usual sense if and only if $\partial\varphi(\mathbf{x})$ consists of a single point, in which case $\partial\varphi(\mathbf{x})=\{\nabla\varphi(\mathbf{x})\}$. See \prettyref{fig:subdiff-schematic} for a one-dimensional example illustrating the subdifferential. 

Next, we define the \emph{Legendre transform} of a general function $\varphi(\mathbf{x})$, which need not be convex. This is the function $\varphi^{*}(\mathbf{p})$ defined for $\mathbf{p}\in\mathbb{R}^2$
by 
\[
\varphi^{*}(\mathbf{p})=\max_{\mathbf{x}\in\mathbb{R}^2}\,\mathbf{x}\cdot \mathbf{p}-\varphi(\mathbf{x}).
\]
Even if a maximizing $\mathbf{x}$ does not exist, the value of $\varphi^{*}(\mathbf{p})$ is still defined (it may be $\infty$). One checks that $\varphi^{*}$ is convex, regardless of the convexity of $\varphi$. If $\varphi$ is convex, its subdifferential and that of its Legendre transform are related in the following  way:
\begin{equation}
\mathbf{p}\in\partial\varphi(\mathbf{x})\quad\iff\quad \mathbf{x}\in\partial\varphi^{*}(\mathbf{p}).\label{eq:abstract-reciprocity}
\end{equation}
Such pairs $(\mathbf{x},\mathbf{p})$ are equivalently characterized by the statement that 
\begin{equation}\label{eq:abstract-reciprocity-2}
\varphi(\mathbf{x}) + \varphi^*(\mathbf{p}) = \mathbf{x}\cdot \mathbf{p}.
\end{equation}
These relations form the backbone of our reciprocal rules for wrinkles, as we shall show.

Finally, we define the \emph{double Legendre transform} of $\varphi$, also know as its \emph{biconjugate}:
\[
\varphi^{**}=(\varphi^{*})^{*}.
\]
For general $\varphi$, which may even take on the value $\infty$, there always holds 
\begin{equation}
\varphi(\mathbf{x})\geq\varphi^{**}(\mathbf{x}).\label{eq:lower-bd-repeatedtransform}
\end{equation}
The function $\varphi^{**}$ is convex, and \prettyref{eq:lower-bd-repeatedtransform}
 ensures its graph lies on or beneath that of $\varphi$. Actually, a stronger statement is true: under certain technical conditions 
 which can be checked to hold in our applications below, $\varphi^{**}$ is the \emph{convex hull} of $\varphi$, i.e., the largest convex function that is never larger than $\varphi$. (In particular, it suffices that the convex hull of $\varphi$ be lower semicontinuous, and that it does not take on the value  $-\infty$.) In symbols,
 \[
 \varphi^{**}(\mathbf{x}) = \max_{\psi}\, \psi (\mathbf{x})
 \]
where the maximization is over all convex functions $\psi$ satisfying $\psi\leq \varphi$. As a result,  $\varphi=\varphi^{**}$ if and only if $\varphi$ is
convex, under the same technical conditions. 

\subsubsection{The relation between $\varphi_{-}$ and $\varphi_{+}$\label{subsubsec:Legendre-transforms}}

We now begin the proof of our simple rules.  The first step is
to realize that the smallest and largest convex extensions $\varphi_{-}$ and $\varphi_{+}$ of $|\mathbf{x}|^2/2$ into any given domain $\Omega\subset\mathbb{R}^2$
are actually Legendre transforms of one another: 
\begin{equation}\label{eq:dual-Legendre-transforms}
\varphi_{+}=\varphi_{-}^{*}\quad\text{and}\quad\varphi_{-}=\varphi_{+}^{*}.
\end{equation}
We shall establish the first part of \prettyref{eq:dual-Legendre-transforms} directly, and then apply the double Legendre transform to recover the second part. 

In \prettyref{subsubsec:abstract-solns} we noted that $\varphi_{+}$ is the convex hull of  
\[
h(\mathbf{x})=\begin{cases}
\infty & \mathbf{x}\in\Omega\\
\frac{1}{2}|\mathbf{x}|^2 & \mathbf{x}\notin\Omega
\end{cases}.
\]
Using the Legendre transform, we now write this as 
\begin{equation}
\varphi_{+}=h^{**}.\label{eq:convexhull-of-h}
\end{equation}
Separately, we observe that 
\begin{equation}
\varphi_{-}=h^{*}.\label{eq:Legendretransform-of-h}
\end{equation}
Indeed, if $\mathbf{x}\in\Omega$ then the shortest distance from it to the boundary satisfies
\[
d_{\partial\Omega}(\mathbf{x})=\min_{\mathbf{y}\in\partial\Omega}\,|\mathbf{x}-\mathbf{y}|=\min_{\mathbf{y}\notin\Omega}\,|\mathbf{x}-\mathbf{y}|,
\]
as the points in $\partial\Omega$ are closer to $\mathbf{x}$ than any other point outside of $\Omega$. By \prettyref{eq:negative-curvature-solution},
\begin{align*}
\varphi_{-}(\mathbf{x}) & =\frac{1}{2}|\mathbf{x}|^{2}-\frac{1}{2}d_{\partial\Omega}^{2}(\mathbf{x}) = \frac{1}{2}|\mathbf{x}|^{2}-\frac{1}{2}\min_{\mathbf{y}\notin\Omega}\,|\mathbf{x}-\mathbf{y}|^{2}\\
&=\max_{\mathbf{y}\notin\Omega}\,\frac{1}{2}|\mathbf{x}|^{2}-\frac{1}{2}\,|\mathbf{x}-\mathbf{y}|^{2} =\max_{\mathbf{y}\notin\Omega}\,\mathbf{x}\cdot \mathbf{y}-\frac{1}{2}|\mathbf{y}|^{2}\\
&=\max_{\mathbf{y}\in\mathbb{R}^2}\,\mathbf{x}\cdot \mathbf{y}-h(\mathbf{y})=h^{*}(\mathbf{x}).
\end{align*}
If $\mathbf{x}\notin\Omega$ then a straightforward manipulation using the definitions, along with the fact that $|\mathbf{x}|^2/2$ is its own Legendre transform, shows that $\varphi_{-}(\mathbf{x})$ and $h^{*}(\mathbf{x})$ are both given by $|\mathbf{x}|^2/2$.

Combining  \prettyref{eq:Legendretransform-of-h}  with \prettyref{eq:convexhull-of-h}, we see that
\[
\varphi_{+}=(h^{*})^{*}=\varphi_{-}^{*}.
\]
Since $\varphi_-$ is convex it follows immediately that
\[
\varphi_{-} = \varphi_{-}^{**} = \varphi_{+}^*.
\]
This proves \prettyref{eq:dual-Legendre-transforms}. For future use, note the following equivalences which are direct consequences of \prettyref{eq:abstract-reciprocity} and \prettyref{eq:abstract-reciprocity-2} along with what we just showed:   
\begin{equation}
\mathbf{p}\in\partial\varphi_{+}(\mathbf{x})\quad\iff\quad \mathbf{x}\in\partial\varphi_{-}(\mathbf{p}) \quad\iff\quad \varphi_+(\mathbf{x}) + \varphi_-(\mathbf{p}) = \mathbf{x}\cdot \mathbf{p}. \label{eq:abstract-reciprocity-two}
\end{equation}

\subsubsection{Identifying $\nabla\varphi_{-}$ and $\nabla\varphi_{+}$\label{subsubsec:Identifying-the-gradients}}

The second step is to show that
\begin{equation}
\nabla\varphi_{-}=\mathbf{P}_{\partial\Omega}\quad\text{and}\quad\nabla\varphi_{+}=\mathbf{P}_{\mathfrak{M}}\label{eq:gradient-formulas}
\end{equation}
almost everywhere in $\Omega$. 
This is probably the hardest step in our proof of \prettyref{eq:formulas-for-sigmaL}. 
First, we clarify the definitions of $\mathbf{P}_{\partial\Omega}$ and $\mathbf{P}_{\mathfrak{M}}$, both of which map $\Omega$ to certain subsets thereof. The reader may wish to consult Fig.~5 in the main text. 

The former map is defined by the condition that $\mathbf{P}_{\partial\Omega}(\mathbf{x})$ is a closest
boundary point to $\mathbf{x}$:
\begin{equation}
\mathbf{P}_{\partial\Omega}(\mathbf{x})\in\partial\Omega\quad\text{and}\quad d_{\partial\Omega}(\mathbf{x})=|\mathbf{x}-\mathbf{P}_{\partial\Omega}(\mathbf{x})|.\label{eq:defn-of-Pbdry}
\end{equation}
Almost every $\mathbf{x}\in \Omega$ admits  a unique closest boundary point. 
The exceptional points are where $\mathbf{P}_{\partial\Omega}$ is not uniquely defined. Such points belong to the \emph{medial axis} of $\Omega$, 
\[
\mathfrak{M} = \left\{\mathbf{p}\in \Omega : d_{\partial\Omega}(\mathbf{p}) = |\mathbf{p}-\mathbf{q}|\text{ for multiple }\mathbf{q}\in\partial\Omega\right\}.
\]
This set appears in white in the $\kappa<0$ shells of Figs.~1-3 of the main text.

Next, we define $\mathbf{P}_{\mathfrak{M}}$. Given $\mathbf{p}\in\Omega$, we first associate to it the set of its closest boundary points,
\[
Q_{\mathbf{p}}=\left\{ \mathbf{q}\in\partial\Omega:d_{\partial\Omega}(\mathbf{p})=|\mathbf{p}-\mathbf{q}|\right\}.
\]
Evidently, $Q_{\mathbf{p}}$ contains two or more points if and only if $\mathbf{p}\in\mathfrak{M}$. 
With this, we let $\mathbf{P}_{\mathfrak{M}}(\mathbf{x})$  be a point on the medial axis whose closest boundary points contain $\mathbf{x}$ in their convex hull:
\begin{equation}\label{eq:defn-of-Pm}
\mathbf{P}_{\mathfrak{M}}(\mathbf{x})\in\mathfrak{M}\quad\text{and}\quad \mathbf{x}\in\text{co}\,Q_{\mathbf{P}_{\mathfrak{M}}(\mathbf{x})}.
\end{equation}
Again, $\mathbf{P}_{\mathfrak{M}}$ is uniquely defined almost everywhere in $\Omega$; in some rare instances $\mathbf{x}$ may belong to multiple convex hulls. Fig.~5a,b in the main text shows a couple typical cases, in which $Q_{\mathbf{p}}$ consists of two and four boundary points, respectively.

We are ready to establish \prettyref{eq:gradient-formulas}. First, recall
from \prettyref{eq:negative-curvature-solution} the formula
\[
\varphi_{-}(\mathbf{x})=\frac{1}{2}|\mathbf{x}|^{2}-\frac{1}{2}d_{\partial\Omega}^{2}(\mathbf{x})
\]
for $\mathbf{x}\in\Omega$. Differentiating almost everywhere yields that 
\[
\nabla\varphi_{-}(\mathbf{x})=\mathbf{x}-d_{\partial\Omega}(\mathbf{x})\nabla d_{\partial\Omega}(\mathbf{x}).
\]
The unit vector $-\nabla d_{\partial\Omega}(\mathbf{x})$
points in the direction of steepest decrease of the shortest distance to the boundary. Hence, $\mathbf{x}-d_{\partial\Omega}\nabla d_{\partial\Omega}(\mathbf{x})$ is a closest boundary point to $\mathbf{x}$. From \prettyref{eq:defn-of-Pbdry}, we get that
\[
\nabla\varphi_{-}(\mathbf{x})=\mathbf{P}_{\partial\Omega}(\mathbf{x})
\]
almost everywhere. This explains the first part of \prettyref{eq:gradient-formulas}.

We turn to $\nabla\varphi_{+}$. This time, we make use of the relationship  between the subdifferentials $\partial\varphi_+$ and $\partial\varphi_-$ obtained at the end of \prettyref{subsubsec:Legendre-transforms}. 
As $\varphi_{+}$ is convex, 
\[
\partial\varphi_{+}(\mathbf{x})=\{\nabla\varphi_{+}(\mathbf{x})\}
\]
for almost every $\mathbf{x}$. For such $\mathbf{x}$, the first equivalence in  \prettyref{eq:abstract-reciprocity-two} simplifies to the statement that
\begin{equation}
\mathbf{x}\in\partial\varphi_{-}(\nabla\varphi_{+}(\mathbf{x}))\label{eq:inclusion_phi_+}.
\end{equation}
We proceed to identify the subdifferential of $\varphi_-$. 

Let $\mathbf{p}\in\Omega$. 
The fact is that $\partial\varphi_{-}(\mathbf{p})$ is the convex hull of the closest boundary points to $\mathbf{p}$, i.e., the set called $\text{co}\,Q_{\mathbf{p}}$. 
To justify this, we turn to the second equivalence in \prettyref{eq:abstract-reciprocity-two}. Using \prettyref{eq:positive-curvature-solution}, we get that $\mathbf{x}\in\partial\varphi_{-}(\mathbf{p})$ if and only if 
\begin{align*}
\varphi_{-}(\mathbf{p}) & =\mathbf{x}\cdot \mathbf{p}-\varphi_{+}(\mathbf{x})=\mathbf{x}\cdot \mathbf{p}-\min_{\sum\theta_{i}\mathbf{q}_{i}=\mathbf{x}}\,\sum\theta_{i}\frac{1}{2}|\mathbf{q}_{i}|^{2}\\
 &= \mathbf{x}\cdot \mathbf{p} - \min_{\sum\theta_{i}\mathbf{q}_{i}=\mathbf{x}}\,\sum\theta_{i} \left ( 
 \frac{1}{2}|\mathbf{q}_{i}-\mathbf{p}|^{2}+ (\mathbf{q}_i-\mathbf{p})\cdot \mathbf{p} + \frac{1}{2}|\mathbf{p}|^2  \right) \\
 & =\frac{1}{2}|\mathbf{p}|^{2}-\frac{1}{2}\min_{\sum\theta_{i}\mathbf{q}_{i}=\mathbf{x}}\,\sum\theta_{i}|\mathbf{q}_{i}-\mathbf{p}|^{2}.
\end{align*}
Each minimization takes place over all finite sets of boundary points $\{\mathbf{q}_{i}\}\subset \partial\Omega$ satisfying
\[
\mathbf{x}=\sum_{i}\theta_{i}\mathbf{q}_{i}\quad\text{where}\quad\sum_{i}\theta_{i}=1\quad\text{and}\quad \theta_{i}\in[0,1].
\] 
Once the $\mathbf{q}_i$ are chosen, the weights $\theta_i$ are determined. Using \prettyref{eq:negative-curvature-solution}, we get that 
\begin{equation}\label{eq:min-me}
\mathbf{x}\in\partial\varphi_{-}(\mathbf{p})\quad\iff\quad d_{\partial\Omega}^{2}(\mathbf{p})=\min_{\sum\theta_{i}\mathbf{q}_{i}=\mathbf{x}}\,\sum\theta_{i}|\mathbf{p}-\mathbf{q}_{i}|^{2}.
\end{equation}
Such $\mathbf{x}$ are apparently convex combinations of optimally chosen boundary points. 
General $\mathbf{q}_i\in\partial\Omega$ satisfy $d_{\partial\Omega}(\mathbf{p})\leq|\mathbf{p}-\mathbf{q}_i|$. Optimal $\mathbf{q}_i$ are therefore characterized by the condition that
\[
d_{\partial\Omega}(\mathbf{p})=|\mathbf{p}-\mathbf{q}_{i}|\quad\text{for each } i.
\] 
We have shown for $\mathbf{p}\in\Omega$ that $\mathbf{x}\in \partial \varphi_{-}(\mathbf{p})$ if and only if $\mathbf{x}$ is a convex combination of closest boundary points to $\mathbf{p}$. Hence, 
\begin{equation}
\partial\varphi_{-}(\mathbf{p})=\text{co}\,Q_{\mathbf{p}}\label{eq:subdifferential-id}
\end{equation}
by the definition of $Q_{\mathbf{p}}$.

Together, \prettyref{eq:inclusion_phi_+} and \prettyref{eq:subdifferential-id} show that
\[
\nabla\varphi_+(\mathbf{x})\in \mathfrak{M} \quad\text{and}\quad \mathbf{x}\in\text{co}\,Q_{\nabla\varphi_{+}(\mathbf{x})}
\]
for almost every $\mathbf{x}\in\Omega$. In particular, the first condition follows from the second one for $\mathbf{x}$ interior to $\Omega$. 
Recall $\mathbf{P}_{\mathfrak{M}}$
was defined in \prettyref{eq:defn-of-Pm} precisely so that these conditions would hold. Putting everything together, we have shown that
\[
\nabla\varphi_{+}(\mathbf{x})=\mathbf{P}_{\mathfrak{M}}(\mathbf{x})
\]
almost everywhere. The proof of \prettyref{eq:gradient-formulas} is complete.

\subsubsection{The simple rules\label{subsubsec:Deriving-the-multipliers}}

We are ready to conclude the formulas for $\boldsymbol{\sigma}_L$ in \prettyref{eq:formulas-for-sigmaL}, and to  establish our simple rules. 
From \prettyref{subsubsec:abstract-solns}, we know that
\begin{equation}
\boldsymbol{\sigma}_L=\begin{cases}
\nabla^{\perp}\nabla^{\perp}\varphi_{-} & \text{if }\kappa\leq 0\\
\nabla^{\perp}\nabla^{\perp}\varphi_{+} & \text{if }\kappa\geq 0
\end{cases}\label{eq:multiplier-formulas-step1}
\end{equation}
where $\varphi_{-}(\mathbf{x})$ and $\varphi_{+}(\mathbf{x})$ are the largest and
smallest convex extensions of $|\mathbf{x}|^2/2$ into $\Omega$.
Their first derivatives were identified in \prettyref{subsubsec:Identifying-the-gradients}, with the result being \prettyref{eq:gradient-formulas}. Those formulas hold almost everywhere in $\Omega$. 
Differentiating once more, we get that 
\[
\nabla\nabla\varphi_{-}=\nabla \mathbf{P}_{\partial\Omega}\quad\text{and}\quad\nabla\nabla\varphi_{+}=\nabla \mathbf{P}_{\mathfrak{M}}
\]
almost everywhere. We proceed to solve for $\boldsymbol{\sigma}_L$.

Recall the notation $\nabla^{\perp}\nabla^{\perp}$ from
\prettyref{eq:grad-grad-notation}.  The required rotations can be done using
\[
\mathbf{R}=\left(\begin{array}{cc}
0 & -1\\
1 & 0
\end{array}\right)
\]
and the observation that
\[
\nabla^{\perp}\nabla^{\perp}=\left(\begin{matrix}\partial_{22} & -\partial_{12}\\
-\partial_{21} & \partial_{11}
\end{matrix}\right)=\mathbf{R}^{T}\left(\begin{matrix}\partial_{11} & \partial_{12}\\
\partial_{21} & \partial_{22}
\end{matrix}\right)\mathbf{R}=\mathbf{R}^{T}\left(\nabla\nabla\right)\mathbf{R}.
\]
Hence,
\begin{equation}
\nabla^{\perp}\nabla^{\perp}\varphi_{-}=\mathbf{R}^{T}\nabla \mathbf{P}_{\partial\Omega}\mathbf{R}\quad\text{and}\quad\nabla^{\perp}\nabla^{\perp}\varphi_{+}=\mathbf{R}^{T}\nabla \mathbf{P}_{\mathfrak{M}}\mathbf{R}.\label{eq:multiplier-formulas-step2}
\end{equation}
 Combining \prettyref{eq:multiplier-formulas-step1} and \prettyref{eq:multiplier-formulas-step2} proves that
\begin{equation}\label{eq:remarkable_formula}
\boldsymbol{\sigma}_L=\begin{cases}
\mathbf{R}^{T}\nabla \mathbf{P}_{\partial\Omega}\mathbf{R} & \text{if }\kappa< 0\\
\mathbf{R}^{T}\nabla \mathbf{P}_{\mathfrak{M}}\mathbf{R} & \text{if }\kappa> 0
\end{cases}
\end{equation}
as originally claimed. 

This remarkably concise formula encodes the mapping from the choice of initial shell to the wrinkle patterns it makes nearby a plane. The link is furnished by the general formula $\boldsymbol{\sigma}_L=\lambda \hat{\mathbf{T}}\otimes\hat{\mathbf{T}}$ obtained in Eq.~(6) of the main text. Recall  $\hat{\mathbf{T}}$ points along the predicted ordered wrinkle peaks and troughs where $\lambda>0$, while in regions where $\lambda=0$ no preferred wrinkle direction is obtained.
The unit vector fields $\hat{\mathbf{T}}_-(\mathbf{x})$ and $\hat{\mathbf{T}}_+(\mathbf{x})$ governing  negatively and positively curved shells are non-null eigenvectors of \prettyref{eq:remarkable_formula}. So,
\[
\hat{\mathbf{T}}_- \times \nabla \mathbf{P}_{\partial\Omega} \neq \mathbf{0}\quad\text{and}\quad \hat{\mathbf{T}}_+ \times \nabla \mathbf{P}_{\mathfrak{M}} \neq \mathbf{0}
\]
and the condition that $\lambda>0$ guarantees that such directions are well-defined. From their definitions at the top of \prettyref{subsubsec:Identifying-the-gradients},  both $\mathbf{P}_{\partial\Omega}(\mathbf{x})$ and $\mathbf{P}_{\mathfrak{M}}(\mathbf{x})$ are constant along at least one line segment through each $\mathbf{x}$ where they are defined. Therefore, $\nabla \mathbf{P}_{\partial\Omega}$ and $\nabla \mathbf{P}_{\mathfrak{M}}$ are at most rank one, almost everywhere. They are also symmetric, as they are rotated versions of $\nabla \nabla \varphi_-$ and $\nabla\nabla \varphi_+$. So, an equivalent characterization of $\hat{\mathbf{T}}_-$ and $\hat{\mathbf{T}}_+$ is that
\begin{equation}\label{eq:stable_dirns_formulas}
\hat{\mathbf{T}}_- \cdot \nabla \mathbf{P}_{\partial\Omega} = \mathbf{0}\quad\text{and}\quad \hat{\mathbf{T}}_+ \cdot \nabla \mathbf{P}_{\mathfrak{M}} = \mathbf{0},
\end{equation}
as claimed in Fig.~5 of the main text. The stated formulas for $\lambda$ now follow.

Rules (i)-(iii) at the start of this section summarize the content of  \prettyref{eq:stable_dirns_formulas}. The first rule asserts that the ordered wrinkles of negatively curved shells follow paths of quickest exit from $\Omega$. Indeed, $\hat{\mathbf{T}}_-(\mathbf{x})$ points along the line segment between $\mathbf{x}$ and $\mathbf{P}_{\partial\Omega}(\mathbf{x})$, since $\mathbf{P}_{\partial\Omega}$ is constant there. 
At the same time, $\hat{\mathbf{T}}_+(\mathbf{x})$  is along the segment between the two closest boundary points to $\mathbf{P}_{\mathfrak{M}}(\mathbf{x})$, in regions where $\nabla \mathbf{P}_{\mathfrak{M}} \neq \mathbf{0}$. This is a special case of \prettyref{eq:defn-of-Pm} combined with \prettyref{eq:stable_dirns_formulas}, so rule (ii) is proved.  
The last rule concerns the disordered regions of positively curved shells. These are captured by the remaining case of \prettyref{eq:defn-of-Pm}, where $\mathbf{P}_{\mathfrak{M}}(\mathbf{x})$ admits three or more closest boundary points. Evidently, $ \mathbf{P}_{\mathfrak{M}}$ is constant throughout their convex hull. So, $\boldsymbol{\sigma}_L=\mathbf{0}$ there, marking the possibility of a disordered response. This completes the proof of our simple rules.